\documentclass[%
reprint,
superscriptaddress,
amsmath,amssymb,
aps,
floatfix]{revtex4-2}

\usepackage{graphicx}% Include figure files
\usepackage{dcolumn}% Align table columns on decimal point
\usepackage{bm}% bold math
\usepackage[hidelinks,colorlinks=true,linkcolor=blue,citecolor=blue,urlcolor=blue]{hyperref}
\usepackage[export]{adjustbox}
\usepackage{xcolor}

\begin{document}
	
	\preprint{Violent relaxation}
	
	\title{Violent relaxation redux}
	
	\author{Michael L. Nastac}
	\email{michael.nastac@physics.ox.ac.uk}
	\affiliation{Rudolf Peierls Centre for Theoretical Physics, University of Oxford, Clarendon Laboratory, Parks Road, Oxford OX1 3PU, UK}
	\affiliation{St. ~John’s College, Oxford OX1 3JP, UK}
	
	\author{Yonadav Barry Ginat}
	\affiliation{Rudolf Peierls Centre for Theoretical Physics, University of Oxford, Clarendon Laboratory, Parks Road, Oxford OX1 3PU, UK}
	
	\author{Robert~J.~Ewart}
	\affiliation{Kavli Institute for Particle Astrophysics and Cosmology, Stanford University, Stanford, CA 94305, USA}
	
	\author{Michael Barnes}
	\affiliation{Rudolf Peierls Centre for Theoretical Physics, University of Oxford, Clarendon Laboratory, Parks Road, Oxford OX1 3PU, UK}
	\affiliation{University College, Oxford OX1 4BH, UK}
	
	\author{Alexander~A.~Schekochihin}
	\affiliation{Rudolf Peierls Centre for Theoretical Physics, University of Oxford, Clarendon Laboratory, Parks Road, Oxford OX1 3PU, UK}
	\affiliation{Merton College, Oxford OX1 4JD, UK}
	
	\date{\today}
	
	\begin{abstract}
	
Collisionless self-gravitating systems, such as merging galaxies, have long seemed an exception to the statistical-mechanical principle that chaotic many-body systems relax to maximum-entropy states: Lynden-Bell's statistical mechanics of their ``violent relaxation,'' constrained by exact conservation of phase-space volume---equivalently, of infinitely many collisionless ``Casimir'' invariants---has appeared not to be a good description. We propose that violent relaxation does give rise to Lynden-Bell's~maximum-entropy~equilibria, but that both are underpinned by \textit{gravitational turbulence}, in which large-scale gravitational fluctuations drive chaotic phase-space mixing of the particles' distribution function. We derive the universal small-scale (sub-Jeans) spectra of this turbulence and show that it leads to phase-space-volume conservation being broken on a timescale only logarithmically longer, in the particle number $N$, than the dynamical (violent-relaxation) time. This timescale separation lets the system relax ``adiabatically'' through a sequence of Lynden-Bell equilibria, each corresponding to instantaneous values of the Casimirs. At long times, it approaches an equilibrium close to its ``ground state'': a distribution whose Casimirs have adjusted so that the system's energy is now its minimum possible energy. Thus, the system's universal tendency is toward a minimum-energy, maximum-entropy state. We corroborate this picture with 1D $N$-body simulations and hypothesize that it can explain the universal equilibria of galaxies and dark-matter halos.

	\end{abstract}
	
	\maketitle
	
	\section*{Introduction}
	
	In 1967, Lynden-Bell \cite{lynden1967statistical} coined the term \textit{violent relaxation} to describe the rapid process by which a collisionless self-gravitating system reaches a virialized (quasi)equilibrium on a few dynamical times—much shorter than the timescale of two-body (collisional) relaxation \cite{chandrasekhar1943dynamical}. Dynamically, violent relaxation is achieved by large-scale, time-dependent gravitational fields that, in the approach to a virial equilibrium, have energy comparable to the kinetic energy of particles and violently rearrange particle orbits in phase space. Lynden-Bell proposed that the end state of violent relaxation could be determined by a statistical-mechanical theory that maximizes entropy (appropriately defined) subject to the fact that a collisionless system obeying the Vlasov equation is phase-space-volume preserving, and, therefore, conserves an infinite family of Casimir invariants \cite{ye1992action}.
	
	Despite its elegance, Lynden-Bell's theory has achieved only limited success in describing the end states of violent relaxation accurately \cite{bindoni2008violent,arad2005inconsistency,arad2005numerical,levin2014nonequilibrium}. This lack  of predictive power has been attributed to the complexity of the process and the diversity of gravitational equilibria observed in nature and numerical simulations \cite{van1982dissipationless, arad2005numerical, arad2005inconsistency, chavanis2006quasi, levin2014nonequilibrium}. Nevertheless, Lynden-Bell's theory has been highly influential, and violent relaxation is thought to underlie structure formation in the Universe—e.g., the formation of dark-matter halos and galaxies \cite{binney2011galactic,mo2010galaxy,fillmore1984self,bertschinger1985self, Peebles1980}. It remains an open problem, however, to explain the mechanisms of violent relaxation and predict its outcomes \cite{bindoni2008violent,mo2010galaxy,arad2005inconsistency}.
	
	In this paper, we propose a conceptual framework to address this problem. Whereas Lynden-Bell assumed exact conservation of phase-space volume during violent relaxation, we argue that its conservation is broken relatively quickly (cf. \cite{e2017arrow}), on a timescale set by the dynamical time multiplied by a logarithmic factor in $N$, the number of particles. The culprit for this fast dissipation is a form of turbulence, which rapidly mixes the system in phase space. We develop a theory of this turbulence and verify it using 1D $N$-body simulations.
	
	In Lynden-Bell’s original theory \cite{lynden1967statistical}, the maximum-entropy state is constrained by the system's initial distribution over the phase-space volume, encoded in an infinite set of Casimirs---the ``waterbag content.'' Instead,  we propose that the system continually strives to relax to a maximum-entropy state, subject to a slowly evolving waterbag content.  The scale separation (in $\log N$) between the dynamical time and the dissipation time allows this relaxation to proceed through a sequence of ``instantaneous'' Lynden-Bell equilibria.
	
	As fluctuations decay, the system strives to approach its ``ground state'': an equilibrium that is both a maximum-entropy (Lynden-Bell) and minimum-energy state. In other words, the principle of the evolution of the waterbag content is to make the system's (conserved) energy asymptotically coincide with the minimum energy that it is allowed to have, given that waterbag content (the end state anticipated in~\cite{wiechen1988relaxation,ziegler1989relaxation}). Our results suggest a promising route to predicting equilibria arising from violent relaxation in more complicated situations, such as galaxy mergers and dark-matter-halo formation in 3D.
	
	\section*{Results}
	
	\subsection*{Vlasov--Poisson system}
	
	We consider an $N$-body system described by the Vlasov--Poisson equations for the single-particle distribution function (DF) $f$ and gravitational potential~$\Phi$:
	\begin{equation} \label{Vlasov}
		\frac{\partial f}{\partial t} + \textbf{v} \cdot \nabla f +\textbf{g} \cdot \nabla_{\mathbf{v}} f = 0,
	\end{equation}
	\begin{equation} \label{Poisson}
		\nabla \cdot \mathbf{g} = -\nabla^2 \Phi = -4 \pi G \rho_{m},
	\end{equation}
	where the gravitational field $\mathbf{g} = -\nabla\Phi$, $G$ is the gravitational constant, and the mass density $\rho_{m} = \int \mathrm{d} \mathbf{v} f$. We formulate our theory for general spatial dimension $d \in \{1,2,3\}$, but focus on $d = 1$ both as a proof of concept and to achieve maximum possible resolution in numerical simulations. The DF is normalized so that $ \iint \mathrm{d} \mathbf{x} \mathrm{d} \mathbf{v} f = Nm = M $, where $m$ is the particle mass (for simplicity, assumed the same for all particles), and $M$ is the total mass. We denote $\rho_{m, 0} = M/L^d$, where $L$ is the characteristic length scale of the system. The constituent particles can be stars \cite{binney2011galactic,hamilton2024kinetic} or dark-matter particles~\cite{rampf2021cosmological}. We have omitted finite-$N$ effects in \eqref{Vlasov}, but we shall take them into account when considering the system's evolution.
	
	\begin{figure*}
		\centering
		\includegraphics[width=\textwidth]{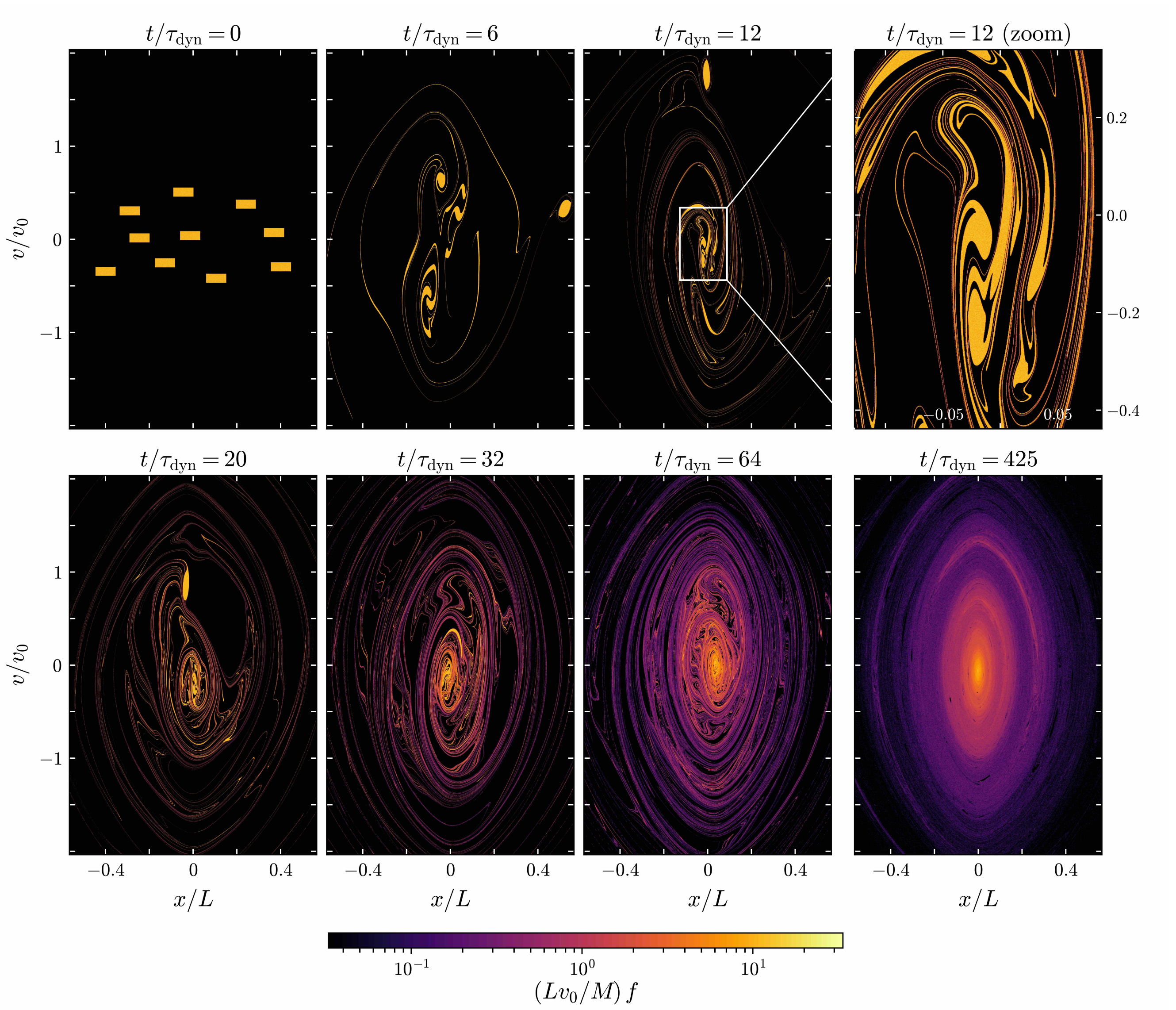}
		\caption{\label{fig:fig1} Contour plots of the DF at a sequence of consecutive times, as indicated, from an $N$-body simulation with $N=10^7$ particles. }
	\end{figure*}

	\subsection*{Violent relaxation}
	
	We consider an isolated, self-gravitating system that is initially far from virial equilibrium and unstable, e.g., due to gravitational collapse or the merging of galaxies. Large-amplitude, rapidly varying fluctuations of the gravitational potential are immediately excited, which drive efficient phase-space mixing on the dynamical timescale~$\tau_{\rm dyn}\sim (G \rho_{m, 0})^{-1/2}$.
	
	This is Lynden-Bell's ``violent relaxation,'' which we simulate in 1D using the $N$-body code \texttt{wendy} \cite{wendy}. We prepare an initial state that is a ``single-waterbag'' distribution in which the DF takes the single value $\eta_0$ on the union of ten equal-sized rectangular patches in phase space with randomly chosen centers, and is zero elsewhere. The initial virial ratio is $2K/U = 0.1$, where
	\begin{align} \label{K_U}
		&K = \iint \mathrm{d} \mathbf{x} \, \mathrm{d} \mathbf{v} \, f(\mathbf{x},\mathbf{v}) \frac{|\mathbf{v}|^2}{2}, \nonumber \\
		& U = \frac{1}{2} \iint \mathrm{d} \mathbf{x} \, \mathrm{d} \mathbf{v} \, f(\mathbf{x},\mathbf{v}) \, \Phi(\mathbf{x})
	\end{align}
	are the kinetic and potential energies of our system, respectively. The total energy is conserved by the system's evolution,~$E=K+U=\mathrm{const}$. We present simulation results in units constructed from the conserved quantities $E$ and $M$ and the gravitational constant $G$: lengths in units of $L = E/ G M^2$, times in units of $\tau_{\rm dyn} = (G\rho_{m, 0})^{-1/2} = E^{1/2}/G M^{3/2}$, and velocities in units of $v_0 = L/\tau_{\rm dyn} = (E/M)^{1/2}$. In these units, the initial phase-space density of the waterbags is~$\eta_0 \simeq 11.3\, M/L v_0$. Further details of the simulations are provided in the Methods section.
	
	Fig.~\ref{fig:fig1} shows contour plots of the DF at successive times. The waterbags undergo gravitational collapse, first assembling into several mini ``halos'' that later merge into a single central halo.
	
\subsection*{Chaotic mixing}

During this evolution, the DF is rotated and stretched in phase space due to linear phase mixing and nonlinear shearing by gravitational fields. These dynamics can be understood by considering the separation of Lagrangian particle trajectories. Consider two nearby particles with phase-space coordinates $(\mathbf{x}_1, \mathbf{v}_1)$ and $(\mathbf{x}_2, \mathbf{v}_2)$. Assuming that the gravitational field is spatially smooth enough to admit a Taylor expansion (this will be shown \textit{a posteriori} to be true), their separation $(\mathbf{r}, \mathbf{u}) \equiv (\mathbf{x}_1-\mathbf{x}_2, \mathbf{v}_1-\mathbf{v}_2)$ satisfies $(\mathrm{d} r^{i}/\mathrm{d} t, \mathrm{d} u^{i}/\mathrm{d} t) = (u^{i}, \sigma^{i}_{j} r^j)$, where $\sigma^{i}_{j} = \partial g^i/\partial x_j = -\partial^2 \Phi/\partial x_i \partial x_j$ is (minus) the Hessian matrix of the gravitational potential, i.e., the tidal matrix, evaluated along the trajectory of the second particle. This linearization holds as long as the separation remains small compared with the outer scale of the gravitational field. Locally in time, depending on the eigenvalues of~$\sigma$, particle separations can either grow exponentially or oscillate along the corresponding eigenvectors, in both position and velocity space (cf.~\cite{nastac2025universal,ginat2025gravitational}). These eigenvalues are constrained by Poisson's equation~\eqref{Poisson}: their sum is $\mathrm{Tr}(\sigma) = -\nabla^2 \Phi = -4 \pi G \rho_{m} \leq 0$. For a distribution of particles, exponential growth and oscillation of particle separations lead to phase-space stretching and rotation of the DF, respectively.

The time dependence of $\Phi$ also leads to stretching and, in 1D, is necessary for it: there, $\sigma$ has a single, non-positive eigenvalue, so there is no instantaneous exponential particle separation. To see how stretching arises instead, decompose $\sigma = \langle\sigma\rangle_T + \tilde\sigma$, where $\langle\sigma\rangle_T = T^{-1}\int_T \mathrm{d}t'\,\sigma(x_2(t'),t') = -4\pi G\langle\rho_m\rangle_T$ is the average of~$\sigma$ along the trajectory of the second particle over a time $T \gg \tau_{\rm dyn}$. The average $\langle\sigma\rangle_T$ rotates particle separations in phase space at frequency $\omega_0 = (4\pi G\langle\rho_m\rangle_T)^{1/2}$; relative to this phase-space rotation, the fluctuations $\tilde\sigma$ then locally shear the DF. When $\Phi$ varies irregularly in space and time, these shears act along irregularly varying directions and, generically, compound to give exponential stretching~\cite{furstenberg1963noncommuting}. Indeed, Fig.~\ref{fig:fig1} shows the DF undergoing a mean rotation in phase space, with filamentary stretching atop it. As the gravitational field evolves on the dynamical timescale $\tau_{\rm dyn}$, this stretching and rotation repeats; a sequence of such events, induced by a gravitational field with spatiotemporal structure, leads to chaotic mixing of the DF in phase space.
	
	\begin{figure*}
		\centering
		\includegraphics[width=\textwidth]{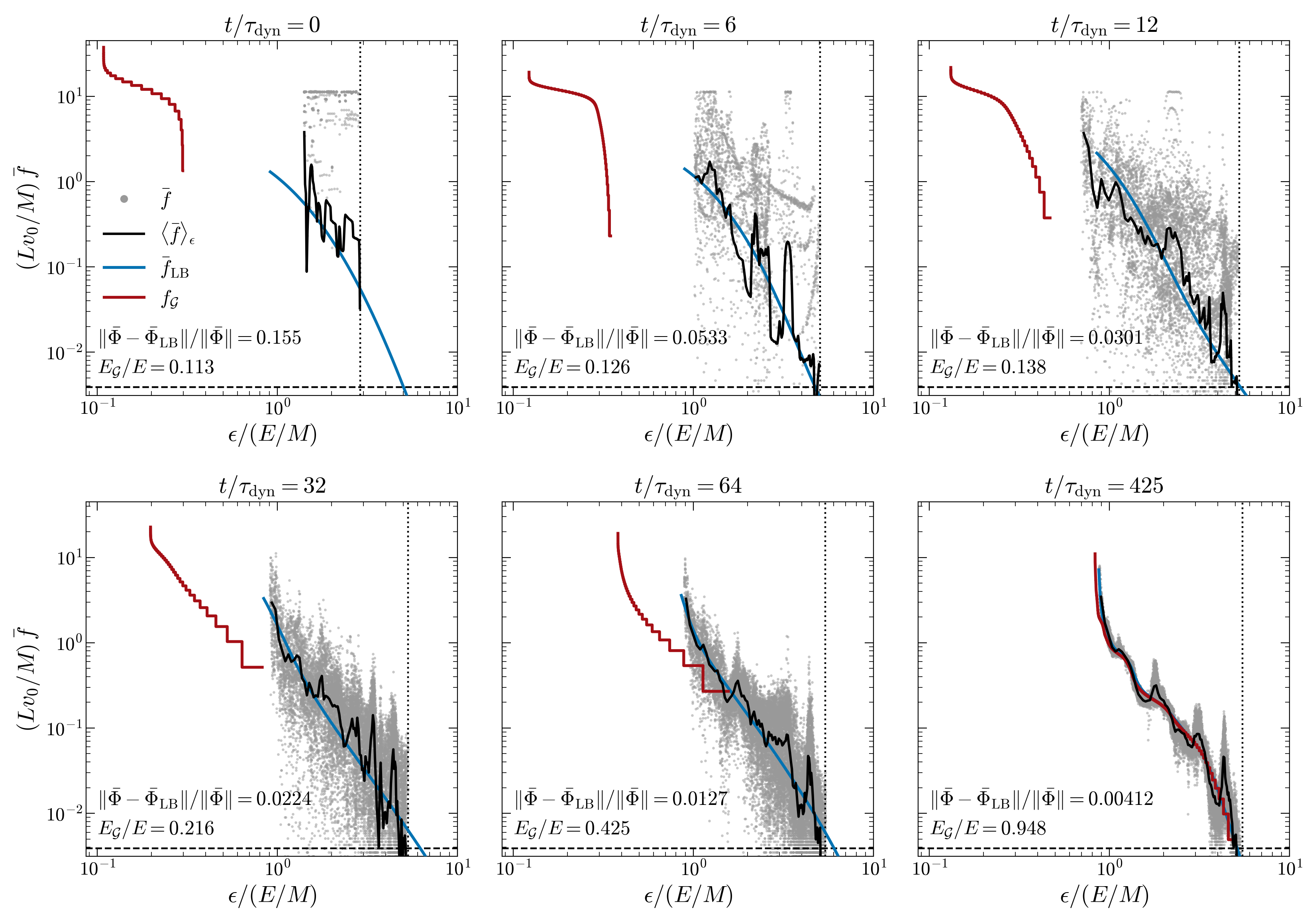}
		\caption{\label{fig:fig3} Scatter plots of the coarse-grained DF $\bar{f}(x, v)$ versus single-particle energy $\epsilon(x,v) = v^2/2 + \bar{\Phi}(x)$ at a sequence of consecutive times, as indicated, from the same $N$-body simulation as in Fig.~\ref{fig:fig1}. The black curve is the energy-averaged DF~\eqref{angle_avg_f}. The blue and red curves are the Lynden-Bell equilibrium \eqref{multi-waterbag_LB} and ground state \eqref{ground_state}, corresponding to the waterbag content at each time, computed using the procedures described in Methods. They are plotted against $\epsilon$ defined with their own self-consistent potentials, $\bar{\Phi}_{\mathrm{LB}}$ and $\Phi_{\mathcal{G}}$, respectively. The horizontal dashed line shows the DF value corresponding to 10 particles per phase-space cell (see Methods for the bin sizes used). The vertical dotted line is the maximum energy per unit mass of any particle at the given time. Legends give the instantaneous values of the ground-state energy $E_{\mathcal{G}}(t)/E$ and of the difference between $\bar{\Phi}_{\mathrm{LB}}$ and $\bar{\Phi}$. The norm $|| \cdot ||$ is the $L^2$ norm.}
	\end{figure*}
	
	\begin{figure}
		\centering
		\includegraphics[width=\linewidth]{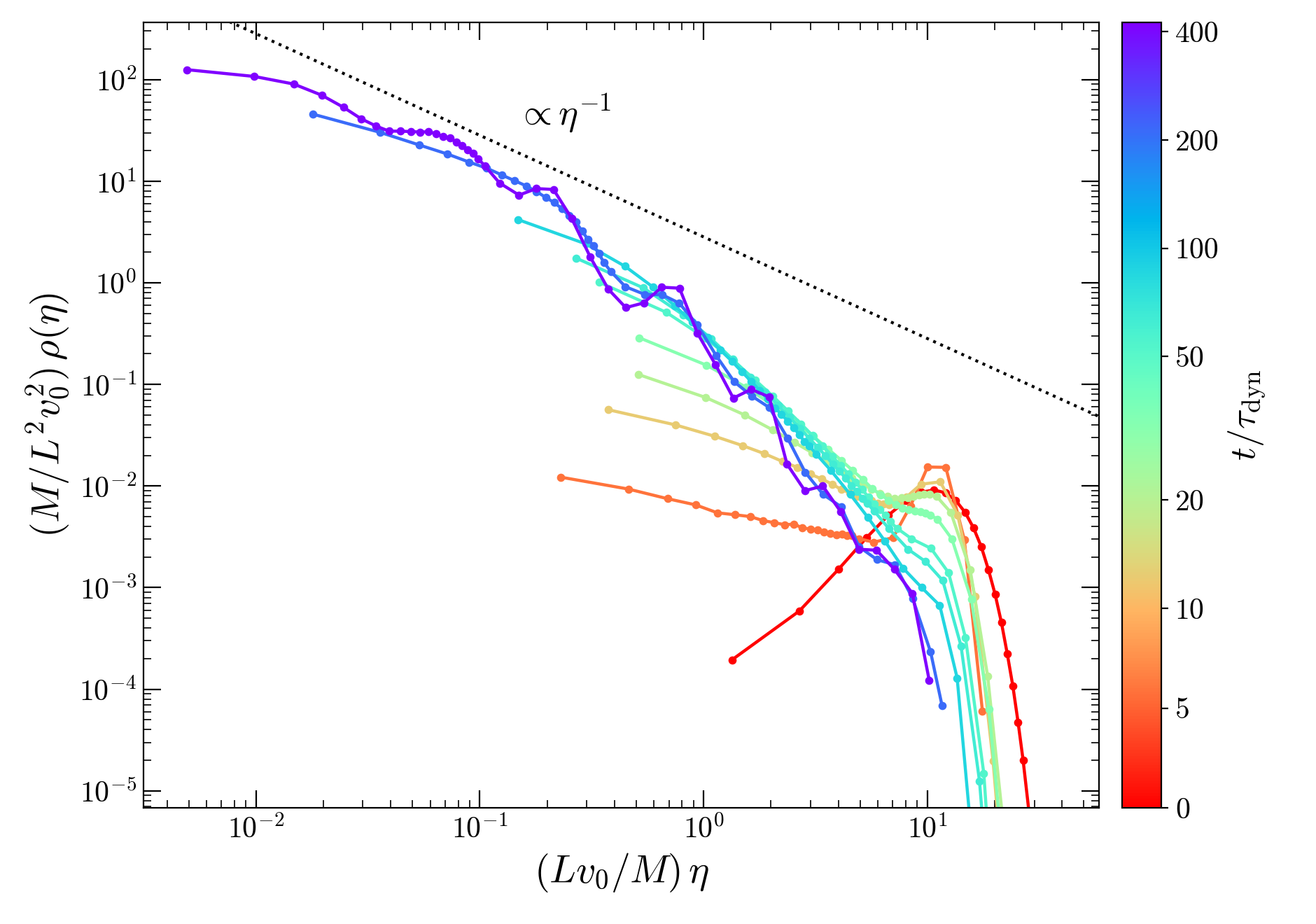}
		\caption{\label{fig:fig4} Waterbag content \eqref{rho_eta_vlasov} of the DF at a sequence of consecutive times (color-coded), as indicated, from the same simulation as in Figs.~\ref{fig:fig1} and~\ref{fig:fig3}. At late times, its low-$\eta$ asymptotic is (marginally) steeper than the $\eta^{-1}$ scaling observed (and well understood) in turbulent, nearly collisionless plasmas \cite{ewart2025relaxation,nastac2025universal}.}
	\end{figure}

	\subsection*{Lynden-Bell equilibria}
	
	Rather than attempt to predict the complex, chaotic structure of the fine-grained DF $f(\mathbf{x},\mathbf{v})$, Lynden-Bell’s statistical mechanics \cite{lynden1967statistical} aims to describe its coarse-grained mean $\bar{f}(\mathbf{x}, \mathbf{v})$ (the same notation is applied to other fields). More generally, it predicts the probability density $P(\mathbf{x},\mathbf{v},\eta)$ that $f(\mathbf{x},\mathbf{v})$ takes the value~$\eta$ at the phase-space point $(\mathbf{x},\mathbf{v})$ \cite{chavanis1996statistical}. Then, by construction,
	\begin{equation} \label{cons_prob}
		\int \mathrm{d} \eta\, P(\mathbf{x},\mathbf{v},\eta) = 1, \qquad
		\bar f(\mathbf{x},\mathbf{v}) \equiv \int \mathrm{d} \eta\, \eta\, P(\mathbf{x},\mathbf{v},\eta).
	\end{equation}
	
	Assuming that phase-space mixing increases disorder, Lynden-Bell posited that the mean distribution maximizes the mixing entropy
	\begin{equation} \label{LB_entropy}
		S_{\mathrm{LB}} = -\int \mathrm{d} \mathbf{x} \mathrm{d} \mathbf{v} \mathrm{d} \eta \, P(\mathbf{x}, \mathbf{v}, \eta) \log P(\mathbf{x}, \mathbf{v}, \eta),
	\end{equation}
	subject to constraints, of which the two most natural ones are the conservation of probability \eqref{cons_prob} and of total energy,
	\begin{equation} \label{energy}
		\iint \mathrm{d} \mathbf{x} \mathrm{d} \mathbf{v} \, \left[\frac{|\mathbf{v}|^2}{2} + \frac{\bar{\Phi}(\mathbf{x})}{2}  \right]   \int \mathrm{d} \eta \, \eta \, P(\mathbf{x}, \mathbf{v}, \eta)  = E
	\end{equation}
(note that linear momentum is trivially conserved in the center-of-mass frame; for $d > 1$, one would also need to consider angular-momentum conservation \cite{lynden1967statistical, varri2012self}). In addition to these invariants, collisionless Vlasov evolution \eqref{Vlasov} conserves the phase-space volume occupied by each level set of $f$. This can be 
	catalogued via the invariance of the ``waterbag content''
	\begin{equation} \label{rho_eta_vlasov}
		\rho(\eta) \equiv \iint \mathrm{d} \mathbf{x}\, \mathrm{d} \mathbf{v} \, \delta\!\left(f(\mathbf{x},\mathbf{v})-\eta\right),
	\end{equation}
	which imposes a continuum of global constraints: for each~$\eta$,
	\begin{equation} \label{rho_eta}
		\iint \mathrm{d}\mathbf{x}\, \mathrm{d} \mathbf{v}\; P(\mathbf{x},\mathbf{v},\eta) = \rho(\eta) = \mathrm{const}.
	\end{equation}
	Equivalently, Vlasov dynamics conserve an infinite family of Casimir invariants \cite{ye1992action}: for any function $\mathcal{H}(f)$ \cite{tremaine1986h},
	\begin{equation}
		\iint \mathrm{d} \mathbf{x}\, \mathrm{d} \mathbf{v} \, \mathcal{H}(f) = \int \mathrm{d} \eta \, \mathcal{H}(\eta) \rho(\eta) = \mathrm{const}.
	\end{equation}
	The conservation of mass follows from the particular case~$\mathcal{H}(f) = f$.
	
	Maximizing \eqref{LB_entropy} subject to all these constraints gives the Lynden-Bell equilibria
	\begin{equation} \label{LB_general}
		P(\mathbf{x}, \mathbf{v}, \eta) = \frac{\mathrm{e}^{-\beta \eta[\epsilon(\mathbf{x}, \mathbf{v})-\mu(\eta)]}}{\int \mathrm{d} \eta' \, \mathrm{e}^{-\beta \eta'[\epsilon(\mathbf{x}, \mathbf{v})-\mu(\eta')]}},
	\end{equation}
where $\beta$ and the ``chemical potential'' $\mu(\eta)$ are the Lagrange multipliers that fix the total energy~\eqref{energy} and the waterbag content~\eqref{rho_eta}, respectively, and~$\epsilon(\mathbf{x}, \mathbf{v}) = |\mathbf{v}|^2/2 + \bar{\Phi}(\mathbf{x})$ is the single-particle energy per unit mass (it contains $\bar\Phi$, not $\bar\Phi/2$ as in \eqref{energy}, because the potential energy is quadratic in $\bar f$, so the factor of $1/2$ cancels upon variation). Here $\bar{\Phi} = \bar{\Phi}_{\mathrm{LB}}$ is the potential generated self-consistently by the equilibrium itself, viz.,
\begin{equation} \label{Poisson_LB}
	\nabla^2 \bar{\Phi}_{\mathrm{LB}} = 4 \pi G \int \mathrm{d} \mathbf{v} \int \mathrm{d} \eta \, \eta \, P(\mathbf{x}, \mathbf{v}, \eta).
\end{equation}
Note that the distribution function \eqref{LB_general} is already explicitly normalized to integrate to unity, as per~\eqref{cons_prob}. 
	
	\subsection*{Single-waterbag Lynden-Bell equilibrium}
	
	For the single-waterbag initial condition in our numerical simulations, $\rho(\eta) \propto\delta(\eta-\eta_0)$. In this case, one obtains the Fermi–Dirac distribution \begin{equation} \label{Fermi_Dirac}
		\bar{f}(\mathbf{x}, \mathbf{v}) = \frac{\eta_0}{1+\mathrm{e}^{\beta \eta_0[\epsilon(\mathbf{x}, \mathbf{v})-\mu_0]}},
	\end{equation}
	reflecting the analogy between Pauli exclusion and phase-space incompressibility: two parts of the distribution occupying different patches of phase space cannot be superimposed by any Vlasov mixing process, just as two fermions cannot be in the same quantum state. In the degenerate limit $\beta\to\infty$, $\bar f\to\eta_0\,H(\mu_0-\epsilon)$, where $H$ is the Heaviside step function. In the non-degenerate limit $\bar{f} \ll \eta_0$,  $\bar f(\mathbf x,\mathbf v)\rightarrow \eta_0\,\mathrm{e}^{-\beta\eta_0[\epsilon(\mathbf x,\mathbf v)-\mu_0]}$, a Maxwellian. Which of these limits is relevant is determined by the relative sparsity of the initial condition within the energetically accessible phase space---in our simulations, the initial state is non-degenerate.
	
	\subsection*{ From initial state to ground state}
	
	Fig.~\ref{fig:fig3} shows scatter plots of $\bar{f}$ versus the particle energy $\epsilon(\mathbf{x},\mathbf{v})$ at a sequence of consecutive times, together with the corresponding mean DF averaged over energy shells,
	\begin{equation} \label{angle_avg_f}
		\langle \bar{f} \rangle_{\epsilon} = \frac{1}{g(\epsilon)} \iint\!\mathrm{d}\mathbf{x}\,\mathrm{d}\mathbf{v} \, \bar{f}(\mathbf{x},\mathbf{v}) \,
		\delta\!\left(\epsilon-\tfrac{|\mathbf{v}|^2}{2}-\bar\Phi(\mathbf{x})\right),
	\end{equation}
	where we have defined the density of states:
	\begin{align}   \label{density_states}
		g(\epsilon) &\equiv \iint\!\mathrm{d}\mathbf{x}\,\mathrm{d}\mathbf{v}\;
		\delta\!\left(\epsilon-\tfrac{|\mathbf{v}|^2}{2}-\bar\Phi(\mathbf{x})\right)
		\nonumber\\
		&\overset{d=1}{=} \sqrt{2}\int_{\bar\Phi(x)<\epsilon}
		\frac{\mathrm{d}x}{\sqrt{\epsilon-\bar\Phi(x)}},
	\end{align}
	with the last equality valid only for $d= 1$.
	
	Since the Lynden-Bell equilibrium corresponding to our initial condition is non-degenerate, it must be close to a Maxwellian. However, the measured $\langle \bar{f} \rangle_{\epsilon}$ is clearly non-Maxwellian, although it does develop an (almost) monotonic profile in $\epsilon$, and the scatter of $\bar{f}$ about this profile tightens with time; i.e., $\bar{f}$ becomes predominantly a function of $\epsilon$. These time-dependent profiles are well approximated by multi-waterbag Lynden-Bell equilibria~\eqref{LB_general},
	\begin{equation} \label{multi-waterbag_LB}
		\bar{f}_{\mathrm{LB}}(\mathbf{x}, \mathbf{v}) = \frac{\int \mathrm{d} \eta \, \eta \, \mathrm{e}^{-\beta \eta[\epsilon(\mathbf{x}, \mathbf{v})-\mu(\eta)]}}{\int \mathrm{d} \eta' \, \mathrm{e}^{-\beta \eta'[\epsilon(\mathbf{x}, \mathbf{v})-\mu(\eta')]}},
	\end{equation}
	but corresponding at each time to the waterbag content~$\rho(\eta)$ of the fine-grained DF measured at that time.
	
	Why the waterbag content evolves rather than remaining constant will be explained momentarily. First, however, we make an observation about the apparent principle guiding its evolution. In the long-time limit, the DF appears to tend toward the system’s ``instantaneous'' \emph{ground state}: the minimum-energy configuration accessible by phase-space-volume-preserving rearrangements at a given $\rho(\eta)$, and so itself time-dependent. In plasma physics, this is known as the Gardner distribution \cite{gardner1963bound}. In self-gravitating systems, it has been referred to as the ``lowest-energy state'' (or ``associated equilibrium'')~\cite{wiechen1988relaxation,aly1989lowest}. Mathematically, the ground state $f_{\mathcal{G}}(\epsilon(\mathbf{x}, \mathbf{v}))$ satisfies, for a given $\rho(\eta)$,
		\begin{align} \label{ground_state}
		\iint \mathrm{d} \mathbf{x}\, \mathrm{d} \mathbf{v} \, \delta\!\left(f_{\mathcal{G}}(\epsilon(\mathbf{x}, \mathbf{v}))-\eta\right) = \rho(\eta),  \\ 		\nabla^2 \Phi_{\mathcal{G}} = 4 \pi G \int \mathrm{d} \mathbf{v} \, f_{\mathcal{G}}(\epsilon(\mathbf{x},\mathbf{v})).
	\end{align}
It can be constructed by ``restacking'' the level sets of~$f$ so that the largest values of $\eta$ occupy the lowest energies (how we compute $f_{\mathcal{G}}$ numerically is described in the Methods). The ground state at each time $t$ is the degenerate ($\beta\to\infty$) limit of the multi-waterbag Lynden-Bell equilibrium \eqref{multi-waterbag_LB}, with energy $E_{\mathcal{G}}(t) \leq E$. This equivalence was proven for collisionless plasmas in \cite{ewart2023nonthermal}, but the same argument extends to collisionless gravitational systems, assuming that the ground state is unique. In homogeneous plasmas, it is known that the~Gardner distribution is unique \cite{dodin2005variational}. For inhomogeneous gravitational systems, no general uniqueness proof, to our knowledge, exists, but we conjecture uniqueness on physical grounds. In Fig.~\ref{fig:fig3}, $f_{\mathcal{G}}$ computed from the simulations is plotted in red.
	
	We stress that the approach to the ground state is \emph{not} due to energy loss—the system is closed, so the total energy is conserved. Rather, the ground state itself evolves, in such a way that the ground-state energy $E_{\mathcal{G}}(t)$ increases toward the system's energy $E$ from below. Since the ground state is determined by the instantaneous waterbag content $\rho(\eta)$, its ability to evolve stems from the breaking of phase-space-volume conservation.
	
	\subsection*{Evolution of waterbag content}
	
	Let us now examine how $\rho(\eta)$ evolves. Fig.~\ref{fig:fig4} shows that it broadens from the initial (approximately) $\delta(\eta-\eta_0)$ to a function with wide support in~$\eta$. The waterbag content changes with time because the phase-space mixing inherent in violent relaxation is a turbulent cascade of the fluctuations of $f$ to ever smaller phase-space scales, generating fine structure down to the scales where fluctuations are dissipated by finite-$N$ effects (particle noise) and the dynamics become irreversible. The timescale over which this dissipation occurs—and hence the timescale for the evolution of $\rho(\eta)$—is
	\begin{equation} \label{tau_dissp}
		\tau_{\mathrm{diss}} \sim  \tau_{\mathrm{dyn}} \log N.
	\end{equation}
	For systems with large $N$, the dissipation time, while longer than $\tau_{\mathrm{dyn}}$, is much shorter than the timescale of collisional two-body relaxation,~$\tau_{\mathrm{b}} \sim \tau_{\mathrm{dyn}} N$~\cite{chandrasekhar1943dynamical}. For example, in our simulations with $N = 10^7$,~$\tau_{\mathrm{diss}}/\tau_{\mathrm{dyn}} \sim 10$, whereas $\tau_{\mathrm{b}}/\tau_{\mathrm{dyn}} \sim 10^7$. We shall derive \eqref{tau_dissp}, and confirm it numerically, later, but first we use it to interpret the observed evolution of our system and propose a new paradigm of violent relaxation.
	
	\subsection*{ Violent relaxation, gentle dissipation, and convergence to ground state}

		\begin{figure}
	\centering
	\includegraphics[width=\linewidth]{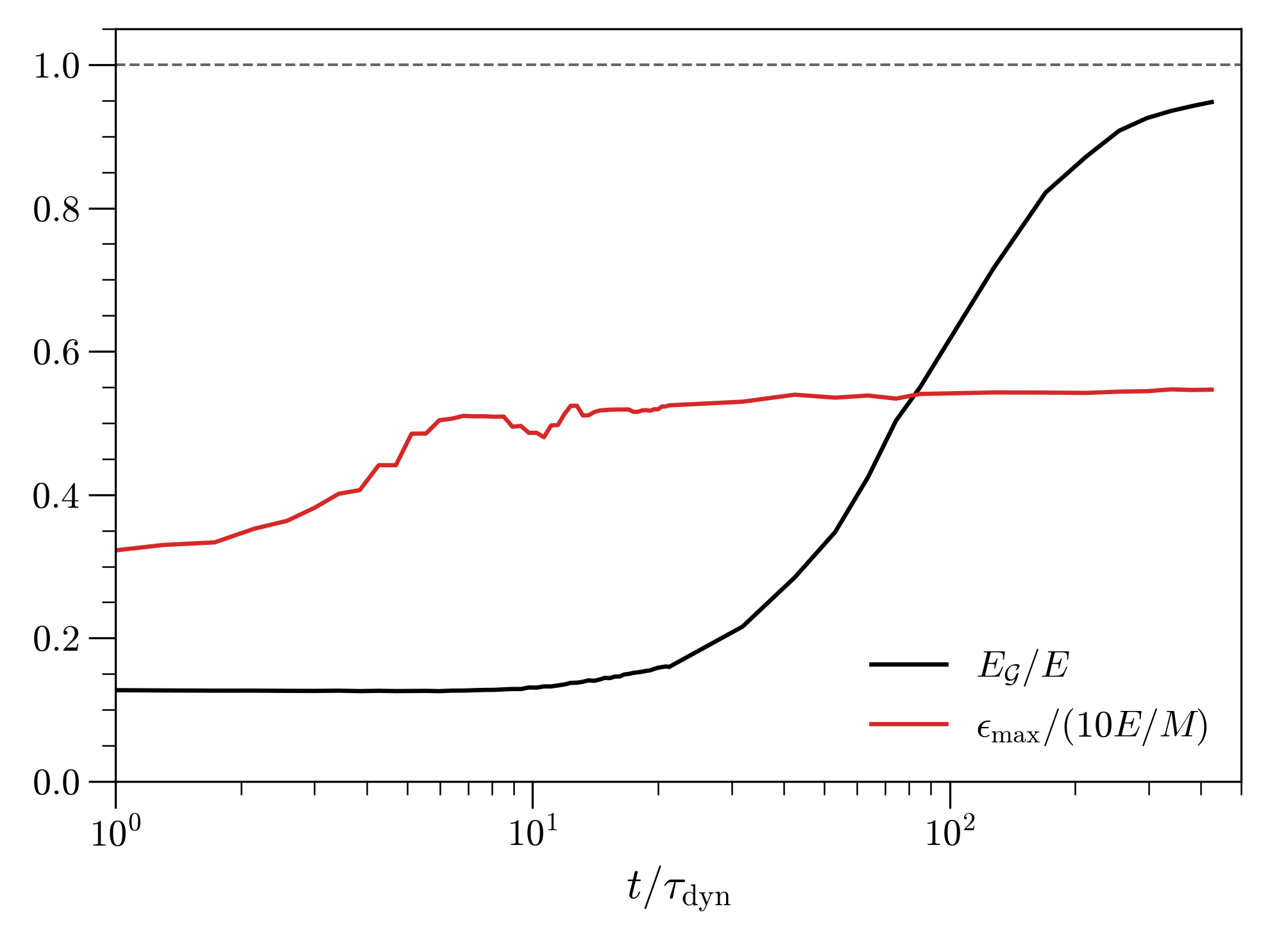}
	\caption{\label{fig:fig_EGE} The black curve shows the ground-state energy $E_{\mathcal{G}}/E$ versus time, from the same simulation as Fig.~\ref{fig:fig1}. The horizontal dashed line, $E_{\mathcal{G}}/E = 1$, corresponds to the fully degenerate limit, in which the system is in its own ground state. The red curve shows the maximum energy per unit mass of any particle versus time (shown as the vertical dotted lines in Fig.~\ref{fig:fig3})---scaled down by a factor of ten for better display. }
\end{figure}
	
	\begin{figure*}
		\centering
		\includegraphics[width=\textwidth]{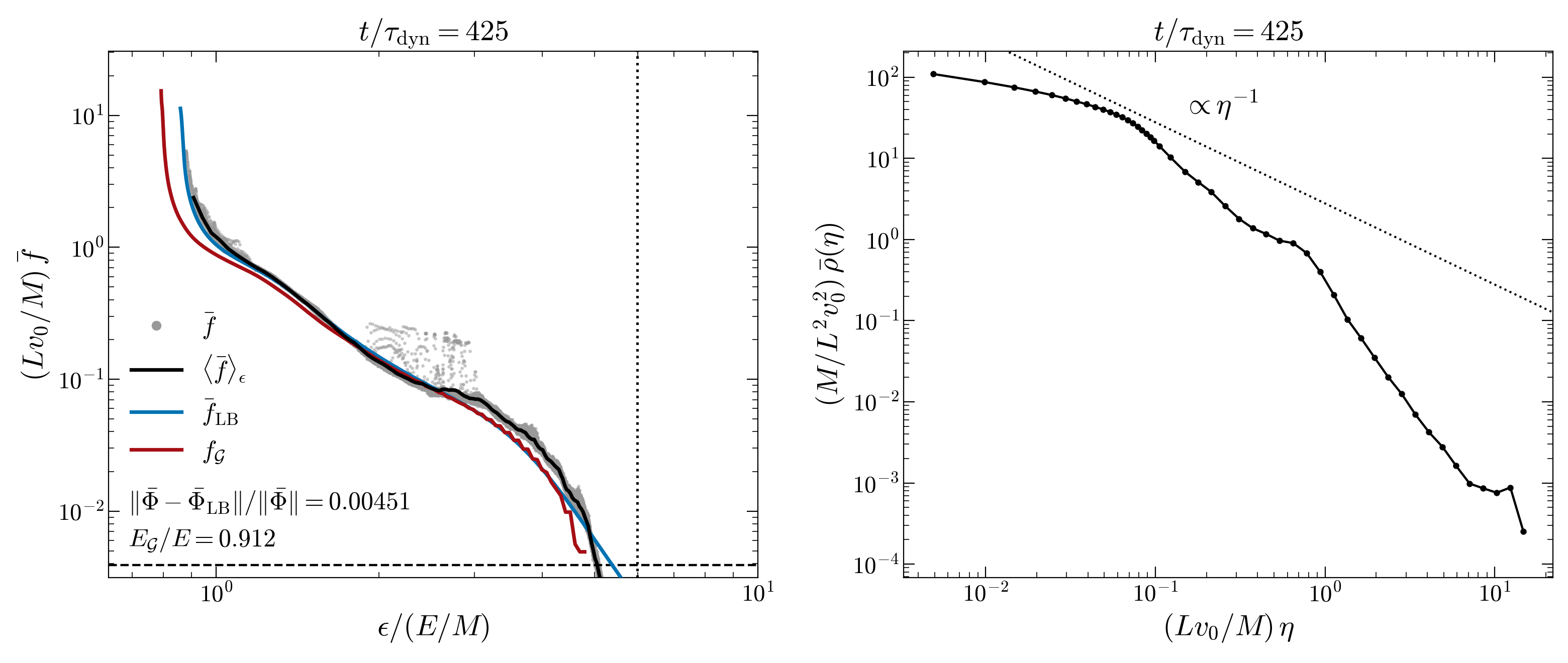}
		\caption{\label{fig:fig_ensemble} Late-time state ($t = 425\,\tau_{\mathrm{dyn}}$) averaged over an ensemble of 100 simulations. Left panel: same as the bottom right panel of Fig.~\ref{fig:fig3} (with various curves having the same meaning), but for the ensemble-averaged DF. Both the Lynden-Bell equilibrium and the ground state are computed from the ensemble-averaged waterbag content. Right panel: the ensemble-averaged waterbag content. The low-$\eta$ asymptotic of $\bar{\rho}(\eta)$ is steeper than the $\eta^{-1}$ scaling found in turbulent, nearly collisionless plasmas \cite{ewart2025relaxation,nastac2025universal} (dotted line; cf.~Fig.~\ref{fig:fig4}).
		}
	\end{figure*}
	
	On the dynamical timescale $\tau_{\mathrm{dyn}}$, the system rapidly mixes and maximizes entropy subject to an approximately conserved waterbag content. On the longer timescale $\tau_{\mathrm{diss}}$, finite-$N$ effects alter $\rho(\eta)$ and progressively erase the system's memory of its earlier states. The system then seeks a new maximum-entropy state subject to the updated waterbag constraint, and the cycle repeats. The $\log N$ separation of timescales, albeit modest, allows the system to relax ``adiabatically'' through a sequence of Lynden-Bell equilibria, each with a fixed ``instantaneous'' waterbag content, but the same total energy and mass set by the initial condition.
		
				In our simulations, the successive equilibria become increasingly degenerate at late times, with $E_{\mathcal{G}}(t)$ approaching $E$ (Fig.~\ref{fig:fig_EGE}). In the scenario that we have proposed, this is inevitable in the long-time limit: as the relaxation proceeds, the turbulent fluctuations $\delta f = f-\bar{f}$ are progressively erased, and if they were removed entirely, $f$ would equal $\bar f_{\mathrm{LB}}$. This is a monotonically decreasing function of energy only, so restacking would leave it unchanged, i.e., the system would be in its ground state. Such a final state---the ground state of the system's own, evolved waterbag content---was envisaged in~\cite{wiechen1988relaxation,ziegler1989relaxation}, with some numerical support from 3D $N$-body simulations~\cite{ziegler1994nonuniform}; here, it emerges as the limit of a sequence of increasingly degenerate Lynden-Bell equilibria.

We stress, however, that the late-time state reached in our simulation is close to, but is not exactly, the system's late-time ground state. At the latest time shown in
Fig.~\ref{fig:fig3}, $E_{\mathcal{G}}/E \approx 0.95$, and the DF retains bumps
and non-monotonic features that persist over many $\tau_{\mathrm{diss}}$. The asymptotic maximum-entropy state has, therefore, not been fully attained: the potential
fluctuations decay before the system can reach it. This is traditionally known
as ``incomplete relaxation'' \cite{lynden1967statistical,chavanis2006quasi}.

Another manifestation of the incompleteness of the relaxation is that the DF has compact
support in energy: at any given time in our simulations, no particle exceeds a maximum energy per unit mass
$\epsilon_{\mathrm{max}}$ (vertical dotted lines in Fig.~\ref{fig:fig3}), and, asymptotically in time, $\epsilon_{\mathrm{max}}$ tends to a finite value (Fig.~\ref{fig:fig_EGE}). The
reason is that, under collisionless dynamics, a particle can gain energy only
while the potential varies in time: along its trajectory,
$\mathrm{d}\epsilon/\mathrm{d}t = \partial\Phi/\partial t$. Since the potential
fluctuations are of order unity only during the early, most violent phase of
the relaxation, the energy that a particle can acquire is bounded. In contrast,
the multi-waterbag Lynden-Bell equilibria \eqref{multi-waterbag_LB} with
$E > E_{\mathcal{G}}$ extend to infinite energies, albeit with exponentially
small tails.

To separate what is universal in the late-time state from what is not, we ran
an ensemble of 100 simulations, each with $N = 10^7$ particles and the same
mass, energy, and initial virial ratio $2K/U = 0.1$, but with different
randomly chosen phase-space positions of the ten initial waterbags (cf.\
Fig.~\ref{fig:fig1}). The ensemble-averaged DF, shown in
Fig.~\ref{fig:fig_ensemble}, is approximately monotonic: the bumps observed in
Fig.~\ref{fig:fig3} are, therefore, realization-dependent (the residual
non-monotonic scatter around $\epsilon \sim 2E/M$ is due to a small
subset of the ensemble, fewer than 10 simulations, in which relatively dense clumps are energized early in the violent relaxation and orbit
outside the core, remaining unmixed). The energy cutoff, in contrast, survives the averaging: a finite $\epsilon_{\mathrm{max}}$ is a dynamical constraint common to all realizations, with its value typically varying by only a few percent across the ensemble. At energies below this cutoff, however, the ensemble-averaged DF is well described by its Lynden-Bell equilibrium and is close to its ground state (Fig.~\ref{fig:fig_ensemble}). Thus, while the relaxation is, strictly speaking, incomplete, it is complete enough for the principle of maximum entropy to hold reasonably well.

	\subsection*{Gravitational turbulence}
	
		Let us now return to the claim that $\rho(\eta)$ evolves on the timescale \eqref{tau_dissp}, due to the turbulent mixing of fluctuations of the phase-space density $f$, and demonstrate that this is indeed what happens.
	
	To motivate gravitational turbulence, we can first consider linear theory. A density fluctuation on scale \(r\) with amplitude \(\delta\rho_{m, r}\) collapses on itself on the dynamical timescale \(\tau_{\mathrm{dyn},r}\sim (G\,\delta\rho_{m, r})^{-1/2}\) if \(\tau_{\mathrm{dyn},r}<\tau_{\mathrm{cross},r}\sim r/v_{\mathrm{th}}\), where $\tau_{\mathrm{cross},r}$ and $v_{\mathrm{th}}$ are the typical crossing time and thermal velocity of the particles, respectively. Physically, the competition between these two timescales reflects the fact that gravitational collapse is stabilized by pressure. The timescales are balanced at the Jeans scale~\cite{jeans1902stability}, $r_{\mathrm{J}} \sim v_{\mathrm{th}} /\sqrt{G \delta \rho_{m, r_{\mathrm{J}}}} \sim v_{\mathrm{th}} \tau_{\mathrm{J}}$. Linear theory then predicts that modes with \(r>r_{\mathrm{J}}\) are Jeans-unstable and undergo collapse, while \(r<r_{\mathrm{J}}\) fluctuations are stable and undergo Landau damping \cite{landau1946vibrations,binney2011galactic}.
	
	In the nonlinear regime relevant for violent relaxation, for scales above the Jeans length, ref.~\cite{ginat2025gravitational} showed---both theoretically and in 1D--1V simulations---that the gravitational collapse of cold dark matter can be understood as a state of phase-space turbulence, in which the DF (and hence also the gravitational field) exhibits a universal fluctuation spectrum in phase space over the range $r_{\mathrm{J}} \ll r \ll r_{\mathrm{nl}}$, where $r_{\mathrm{nl}}$ is the characteristic scale at which nonlinearity becomes important (cf. \cite{konrad2022asymptotic}).
	
	In this paper, we focus on fluctuations below the Jeans scale. This is because dissipation is achieved by particle noise erasing fluctuations on the smallest scales in the system. Fluctuations generated above the Jeans scale inevitably cascade down to it (via a sequence of Jeans collapses), and then, by a somewhat different mechanism, to scales below it, before ultimately being dissipated: submerged below the noise floor.
	
	\subsection*{Cascaded invariant}
	
	To build a turbulence theory, we need an invariant that is cascaded from large to small scales. We posit that the cascaded invariant is \cite{diamond2010modern, nastac2023phase,nastac2025universal,ginat2025gravitational}
	\begin{equation} \label{C2}
		C_2 = \frac{1}{V} \iint \mathrm{d} \mathbf{x} \mathrm{d} \mathbf{v} \, f^2,
	\end{equation}
	where $V$ is the system's spatial volume, taken to be the volume of the smallest $d$-dimensional cube that encloses all bound particles throughout the evolution. $C_2$  is known as an H-function in galactic dynamics \cite{tremaine1986h} and a Casimir invariant (or, ``enstrophy'' \cite{servidio2017magnetospheric, knorr1977time}, ``phasestrophy'' \cite{lesur2013nonlinear, diamond2010modern}, or ``$f$-strophy'' \cite{schekochihin2022lectures}) in plasma physics~\cite{ye1992action}. $C_2$ is conserved by collisionless evolution and dissipated by diffusion or phase-space coarse-graining, so it is a generalized (negative) entropy. It also satisfies Parseval's theorem: $\overline{C_2} = V/(2 \pi)^{2d} \iint \mathrm{d} \mathbf{k} \mathrm{d} \mathbf{s} P_{f}(\mathbf{k}, \mathbf{s})$, where $P_{f}(\mathbf{k}, \mathbf{s}) = \overline{|\hat{f}(\mathbf{k}, \mathbf{s})|^2}$ is the fluctuation spectrum, $\mathbf{k}$ is a spatial wavevector, $\mathbf{s}$ is a velocity-space wavevector, and $\hat{f}(\mathbf{k}, \mathbf{s}) = V^{-1} \iint \mathrm{d}\mathbf{x}\, \mathrm{d}\mathbf{v}\, \mathrm{e}^{-\mathrm{i}(\mathbf{k}\cdot\mathbf{x} - \mathbf{s}\cdot\mathbf{v})} f(\mathbf{x}, \mathbf{v})$ is the Fourier transform of $f(\mathbf{x}, \mathbf{v})$ in both arguments. $C_2$ is, therefore, a quantity that can be thought of as transferred from large to small scales (equivalently, from small to large wavenumbers) when the DF undergoes mixing. Since the spectral scalings that we derive hold only in a statistical sense, here and below we take the overline to denote an ensemble average (alternatively, one could, e.g., take a time average over several dynamical times).
	
	In addition to the above properties, note that~$P_{f}(\mathbf{k}, \mathbf{0}) = P_{\rho_m}(\mathbf{k}) = \overline{|\hat{\rho}_m(\mathbf{k})|^2}$ \cite{nastac2025universal,ginat2025gravitational}. By Poisson's equation \eqref{Poisson}, the fluctuation spectrum of the mass density is directly related to the fluctuation spectrum of the gravitational field: $P_{\mathbf{g}}(\mathbf{k}) = 16 \pi^2 G^2 P_{\rho_m}(\mathbf{k})/|\mathbf{k}|^2$, where $P_{\mathbf{g}}(\mathbf{k}) = \overline{|\hat{\mathbf{g}}(\mathbf{k})|^2}$. Thus, to quantify phase-space mixing and determine the fluctuation spectrum of the gravitational field, we need $P_{f}(\mathbf{k}, \mathbf{s})$ as a function of both of its phase-space variables.
	
	\subsection*{Fluctuation spectra}
	
		\begin{figure}
		\centering
		\includegraphics[width=\linewidth]{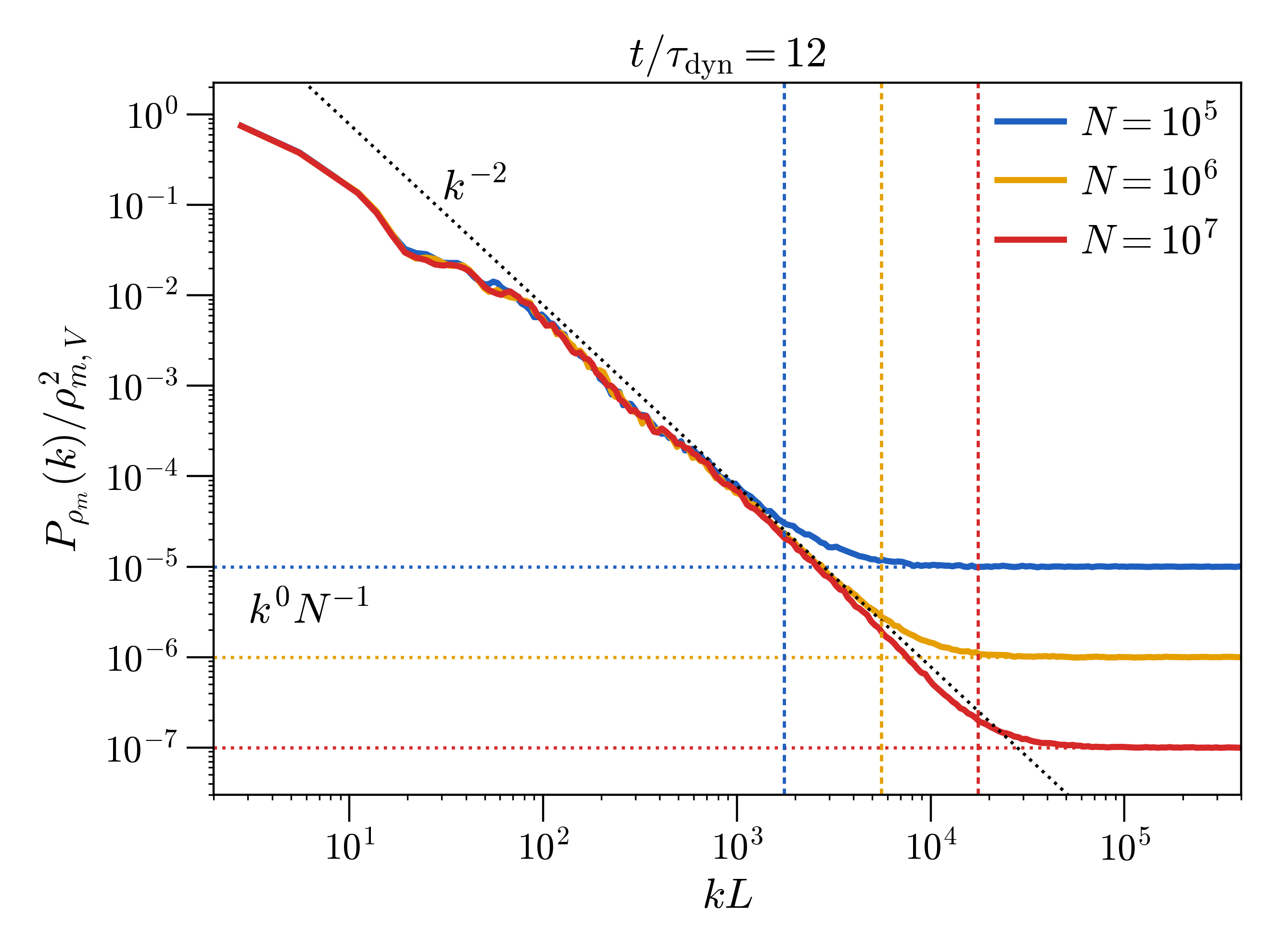}
		\caption{\label{fig:fig5} Ensemble-averaged mass-density fluctuation spectrum, computed at a time when the mixing is strong (after at least one $\tau_{\mathrm{diss}}$ for each $N$). Here $\rho_{m,V} = M/V$ is the mean mass density in the spatial domain of size $V$ over which the Fourier transform is taken [see \eqref{C2}]. The spectrum is computed from ensembles of 100 simulations each with $N = 10^5$ (blue), $10^6$ (orange), and $10^7$ (red) particles (for the $N = 10^7$ case, the snapshot $t = 12\,\tau_{\rm dyn}$ shown in Fig.~\ref{fig:fig1} is a member of the ensemble used). The horizontal dotted lines show the noise floor corresponding to each $N$, according to \eqref{noise_spec}. The red vertical line corresponds to the wavenumber at which the measured density spectrum for the $N=10^7$ ensemble approaches the noise floor within a factor of two---the other two vertical lines are scaled from this line according to \eqref{k_noise}.}
	\end{figure}
	
	\begin{figure*}[t]
		\centering
		\includegraphics[width=\textwidth]{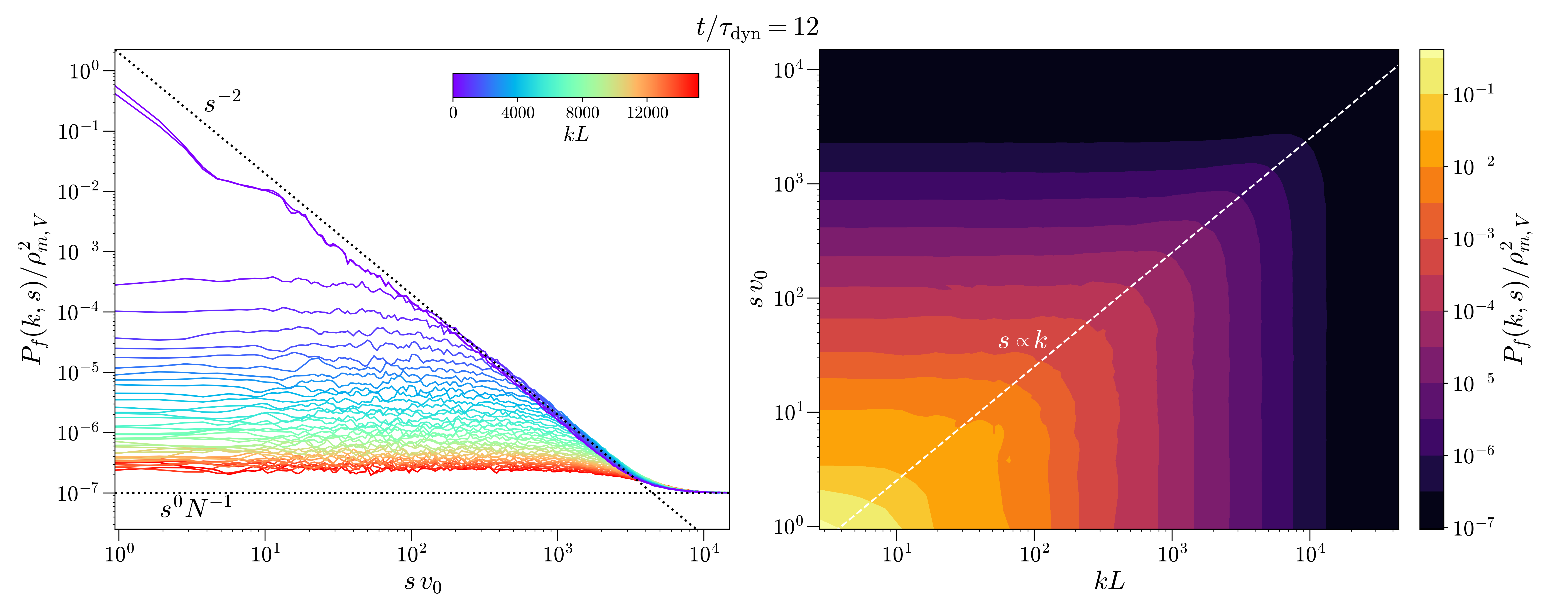}
		\caption{\label{fig:fig6}
			Phase-space spectrum $P_f(k,s)$ at $t = 12\,\tau_{\rm dyn}$ (the same time as in Fig.~\ref{fig:fig5}), ensemble-averaged over 100 simulations with $N = 10^7$ particles. Left panel: 1D cuts of $P_f(k,s)$ as a function of $s > 0$, for a sequence of values of $k$ (color bar). Right panel: the full 2D spectrum. The white dashed line is the critical-balance line, $s \sim \gamma^{-1} k$, separating the regions where the nonlinear mixing dominates ($s \gg \gamma^{-1} k$) and those where the linear phase mixing does ($s \ll \gamma^{-1} k$).}
	\end{figure*}
	
	In the Methods, we formulate a Kolmogorov-style scaling theory for sub-Jeans-scale gravitational turbulence. The result is that there exists an inertial range in phase space in which the spectrum of $C_2$ scales as
	\begin{equation} \label{F_ks_solved}
		\langle P_{ f} \rangle_{k,s} \sim \varepsilon 
		\begin{cases} 
			k^{d-1} \left(\gamma s\right)^{-d-1} , & k \ll \gamma s, \\
			\left(\gamma s \right)^{d-1} k^{-d-1} , & k \gg \gamma s,
		\end{cases}
	\end{equation}
	where $k = |\mathbf{k}|$, $s = |\mathbf{s}|$, and $\langle ...\rangle_{k,s}$ denotes shell averaging in wavenumber space, i.e., over the directions of $\mathbf{k}$ and $\mathbf{s}$, including the factor $\sim V k^{d-1} s^{d-1}$. The spectrum results from a phase-space cascade of $C_2$ across both position and velocity space, driven by the combined action of linear phase mixing and nonlinear gravitational mixing---i.e., the ``random'' stretching and rotation of the DF described earlier, at every scale. The associated linear and nonlinear timescales are in ``critical balance'' at every scale  \cite{nastac2023phase, nastac2025universal, ginat2025gravitational}, yielding a scale-independent cascade rate $\gamma$ and a self-similar spectrum whose structure is organized along the critical-balance line $k \sim \gamma s$.
	
	The mass-density fluctuation spectrum can be found by taking the $s \rightarrow 0$ limit of the spectrum \eqref{F_ks_solved} not shell-averaged in $s$, i.e., with a factor of $s^{d-1}$ removed:
	\begin{equation} \label{density_spec}
		\langle P_{\rho_m}\rangle_{k} \sim \varepsilon \gamma^{d-1} k^{-d-1}.
	\end{equation}
	By combining \eqref{density_spec} with \eqref{Poisson}, we can also  obtain the fluctuation spectrum for the gravitational field:
	\begin{equation} \label{gk}
		\langle P_{\mathbf{g}}\rangle_{k}  \sim \varepsilon \, G^2 \gamma^{d-1} \, k^{-d-3}.
	\end{equation}
	These are the universal fluctuation spectra of Vlasov--Poisson gravitational turbulence below the Jeans scale. The fact that \eqref{gk} is steeper than $k^{-3}$ in any $d$ implies that the mixing of the DF at small scales is dominated by the gravitational field at the outer scale of the turbulence, consistent with the cascade rate being independent of scale \cite{frisch1995turbulence}. Such non-local interactions make this turbulence resemble passive-scalar mixing in the Batchelor regime \cite{batchelor1959small}, and, even more closely, the Vlasov--Poisson plasma turbulence below the Debye length \cite{nastac2025universal}. Indeed, up to physical constants, \eqref{F_ks_solved} and \eqref{gk} are the same as the fluctuation spectra of the DF and electric field in sub-Debye-scale, collisionless plasma turbulence \cite{nastac2025universal}. This correspondence arises because the underlying physical mechanism of the cascade—phase-space mixing driven by the interplay between linear phase mixing and nonlinear acceleration by large-scale fields—is the same in both systems.
	
	Fig.~\ref{fig:fig5} shows the density power spectrum from our simulations, which indeed agrees well with the predicted $k^{-2}$ scaling. In Fig.~\ref{fig:fig6}, we also confirm the phase-space structure predicted by \eqref{F_ks_solved}, including its $s^{-2}$ scaling at high~$s$. The same scalings below the Jeans scale have been reported in 1D continuum Vlasov--Poisson simulations of gravitationally collapsing cold dark-matter streams \cite{ginat2025gravitational}, consistent with our expectation that the sub-Jeans fluctuation spectrum \eqref{F_ks_solved} is a universal feature of violent relaxation, independent of the initial conditions and of the details of the large-scale dynamics.

	\subsection*{Cascade rate}
	
	By \eqref{Poisson}, $\delta g_r \sim G \delta \rho_{m, r} r$. Evaluating this expression at the Jeans scale, together with the fact that $\tau_{\mathrm{J}} \sim 1/\sqrt{G \delta \rho_{m, r_{\mathrm{J}}}}$, yields $\delta g_{r_{\mathrm{J}}} \sim \tau_{\mathrm{J}}^{-2} \, r_{\mathrm{J}}$. Thus, $\gamma \sim \tau_{\mathrm{J}}^{-1} $, viz., the cascade time is the dynamical time of gravitational collapse. For order-unity fluctuations, $\delta \rho_{m, r_{\mathrm{J}}} \sim \rho_{m, 0}$, we have $\tau_{\mathrm{J}} \sim \tau_{\mathrm{dyn}}$. This is consistent with the standard notion that violent relaxation occurs on a characteristic timescale of the order of the dynamical time \cite{lynden1967statistical,kandrup1990fast}.
	
	\subsection*{Outer scale}
	
	In position space, the cascade starts at the Jeans scale~$r_{\mathrm{J}}$: any larger scale is unstable to gravitational collapse and falls under the regime studied in \cite{ginat2025gravitational}. Assuming the fluctuations are critically balanced at the Jeans scale, the largest velocity-space scale of the cascade is $r_{\mathrm{J}}/\tau_{\mathrm{J}} \sim v_{\mathrm{th}}$.
	
	Critical balance at the outer (Jeans) scale implies that the gravitational-energy density is comparable to the kinetic-energy density: $\delta g^2_{r_{\mathrm{J}}}/G \sim \tau_{\mathrm{J}}^{-4} \, r^2_{\mathrm{J}}/G \sim \delta \rho_{m, r_{\mathrm{J}}} v_{\mathrm{th}}^2$. This reflects the tendency of violent relaxation to drive the system toward virial equilibrium, in which gravitational and kinetic energies are in approximate balance. The rapid temporal variation of the gravitational field's fluctuations—on the timescale $\gamma^{-1} \sim \tau_{\mathrm{J}}$—combined with their large amplitudes, changes particle energies by an order-unity amount on a dynamical timescale. This is the sense in which the relaxation is ``violent,'' in agreement with Lynden-Bell’s original vision of the process~\cite{lynden1967statistical}.
	
	\subsection*{Dissipation cutoff}
	
	In the absence of turbulence, a gravitational system will have fluctuations due to discrete particle noise. For $N$ uncorrelated point particles, the (shell-averaged) spectra of their noise are \cite{nastac2025universal,klimontovich2013statistical}
	
	\begin{align} \label{noise_spec}
		&\langle P_{ f} \rangle_{k,s} \sim \frac{M^2}{N V} k^{d-1}s^{d-1}, \, 	\langle P_{ \rho_m} \rangle_{k} \sim \frac{M^2}{N V} k^{d-1}, \nonumber \\ & \langle P_{\mathbf{g}}\rangle_{k} \sim \frac{ G^2 M^2}{N V} k^{d-3}.
	\end{align}
	In order for the turbulence to exist, its spectrum must be above this noise floor. The turbulence spectrum \eqref{gk} meets the noise spectrum \eqref{noise_spec} at the wavenumber
	\begin{equation} \label{k_noise}
		k_{\mathrm{diss}}r_{\mathrm{J}} \sim N_{\mathrm{J}}^{1/2d} \left(\frac{\delta \rho_{m, r_{\mathrm{J}}}}{\rho_{m, 0}}\right)^{1/d},
	\end{equation}
	where $N_{\mathrm{J}} \equiv N r^d_{\mathrm{J}}/V$ is the number of particles in a sphere of radius $r_{\mathrm{J}}$, and we have estimated $\varepsilon \sim \delta \rho_{m, r_{\mathrm{J}}}^2/v_{\mathrm{th}}^d \tau_{\mathrm{J}}$. For $N_{\mathrm{J}} \gg 1$ and large-amplitude fluctuations, $\delta \rho_{m, r_\mathrm{J}} \sim \rho_{m, 0}$, \eqref{k_noise} satisfies $k_{\mathrm{diss}}r_{\mathrm{J}} \gg 1$, so the turbulence has the inertial range $r_{\mathrm{diss}} \ll r \ll r_{\mathrm{J}}$, where $r_{\mathrm{diss}} \sim k_{\mathrm{diss}}^{-1}$.
	
	Fig.~\ref{fig:fig5} shows that the density spectrum is indeed submerged into the noise floor $\propto k^0$ at high $k$, and increasing $N$ widens the inertial range of the cascade, consistent with the scaling \eqref{k_noise}.
	
	By critical balance, the largest wavenumber of the turbulence in velocity space is $s_{\mathrm{diss}} \sim \tau_{\mathrm{J}} \, k_{\mathrm{diss}}$, which satisfies $s_{\mathrm{diss}} v_{\mathrm{th}} \sim k_{\mathrm{diss}} r_{\mathrm{J}}$. Therefore, when $N_{\mathrm{J}} \gg 1$ and $\delta \rho_{m, r_\mathrm{J}} \sim \rho_{m, 0}$, the turbulence has the velocity-space inertial range $u_{\mathrm{diss}} \ll u \ll v_{\mathrm{th}}$, where $u_{\mathrm{diss}} \sim s_{\mathrm{diss}}^{-1}$.

	\subsection*{Dissipation time}
	
	Thus, during violent relaxation, fluctuations are mixed in phase space down to the scale $(r_{\mathrm{diss}}, u_{\mathrm{diss}})$, below which turbulent fluctuations are superseded by a noise floor extending all the way down to the scale of the interparticle spacing. The scale $(r_{\mathrm{diss}}, u_{\mathrm{diss}})$ is a natural coarse-graining scale for the DF: fluctuations on larger scales are incorporated into the smooth, coarse-grained distribution, while those on smaller scales are considered dissipated (cf. \cite{nastac2025universal}).
	
\begin{figure}
		\centering
		\includegraphics[width=\linewidth]{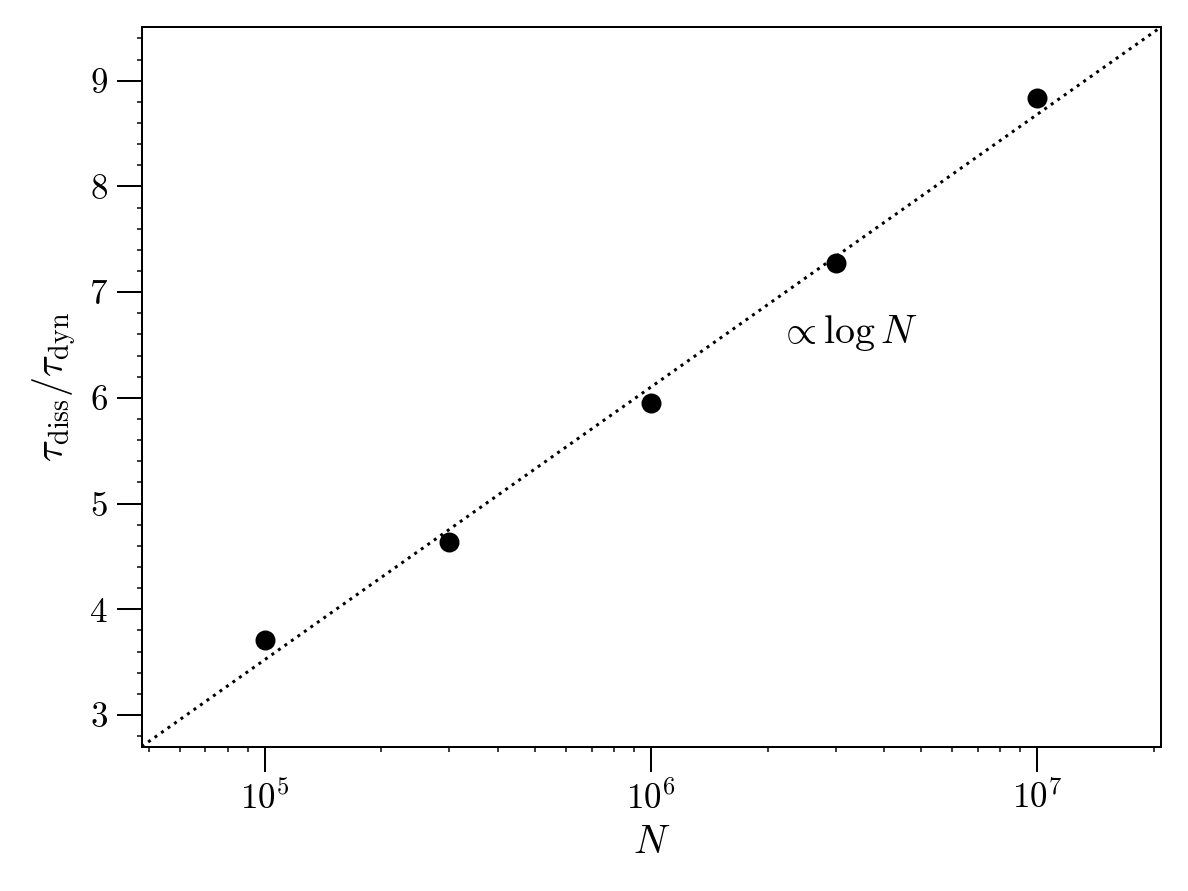}
		\caption{\label{fig:fig7} Dissipation time $\tau_{\mathrm{diss}}$ versus $N$. The dotted line is~$\propto \log N$. The procedure used to estimate $\tau_{\mathrm{diss}}$ is described in Methods.}
	\end{figure}

	Let us compute the characteristic timescale $\tau_{\mathrm{diss}}$ on which turbulent fluctuations are processed from the Jeans scale down to the dissipation scale. We can estimate $\tau_{\mathrm{diss}}$ by dividing the total $C_2$ stored in phase space by its injection rate $\varepsilon$. Integrating \eqref{F_ks_solved} over $s$ yields $\int \mathrm{d} s \langle P_{ f} \rangle_{k,s} \sim \varepsilon \gamma^{-1} k^{-1}$. Further integrating this 1D $k$ spectrum from $r_{\mathrm{J}}^{-1}$ up to $k_{\mathrm{diss}}$, and using $\gamma \sim \tau_{\mathrm{J}}^{-1}$, gives $\overline{C_2} \sim \varepsilon \, \tau_{\mathrm{J}} \log (k_{\mathrm{diss}}r_{\mathrm{J}})$. Therefore, using \eqref{k_noise},
	\begin{equation} \label{tau_dissp_from_C2}
		\tau_{\mathrm{diss}} \sim \frac{\overline{C_2}}{\varepsilon} \sim  \tau_{\mathrm{J}} \log \left[ N_{\mathrm{J}}^{1/2d} \left( \frac{\delta \rho_{m, r_\mathrm{J}}}{\rho_{m, 0}} \right)^{1/d} \right].
	\end{equation}
	For order-unity fluctuations, $\delta \rho_{m, r_\mathrm{J}}\sim \rho_{m, 0}$, we have $\tau_{\mathrm{J}}\sim \tau_{\mathrm{dyn}}$, so \eqref{tau_dissp_from_C2} reduces to \eqref{tau_dissp}. In Fig.~\ref{fig:fig7}, we verify this $\log N$ scaling numerically. Although we only explicitly estimated the dissipation timescale of $C_2$, we expect the other Casimirs to be broken, and hence $\rho(\eta)$ to evolve, on the same timescale.
	
	The timescale \eqref{tau_dissp} is short from a conventional viewpoint—conservation of phase-space volume is broken in finite-$N$ systems far sooner than after a two-body (binary-collision) relaxation time $\tau_b \sim \tau_{\mathrm{dyn}} N$—yet $\tau_{\mathrm{diss}}$ is long compared to the dynamical time. This separation, while modest, is crucial to the paradigm of violent relaxation that we advocate in this paper. It allows the system to keep maximizing entropy and reaching ``instantaneous'' Lynden-Bell equilibria faster than its memory, encoded in $\rho(\eta)$, is scrambled by dissipation.
	
	\section*{Discussion}
	
	We have proposed a new conceptual framework for violent relaxation, in which collisionless self-gravitating systems maximize the mixing entropy \eqref{LB_entropy} on dynamical timescales, subject to Casimir constraints, encoded in the waterbag content $\rho(\eta)$ \eqref{rho_eta_vlasov}, which evolves on the logarithmically longer dissipation timescale \eqref{tau_dissp}, determined by the way in which turbulent mixing occurs in phase space. The separation between the two timescales enables relaxation through a sequence of Lynden-Bell equilibria. In the long-time limit, the system strives to reach a simultaneously maximum-entropy (Lynden-Bell) and minimum-energy (Gardner) state, meaning that its Casimirs evolve in such a way that its ground state subject to these constraints has an energy that asymptotes to the system's energy from below.
	
The natural direction of travel (and ambitious hope) is to use this framework to explain the universal equilibria of galaxies and dark-matter halos \cite{binney2011galactic,mo2010galaxy}, such as the Navarro–Frenk–White density profile \cite{navarro1997universal}. In our picture, universality would arise from universal attractors for $\rho(\eta)$. Recent work \cite{ewart2025relaxation} indicates that a low-$\eta$ asymptotic $\rho(\eta) \propto \eta^{-1}$ is universal in collisionless plasma turbulence; by contrast, our simulations of a relaxing gravitational system show a steeper scaling (Figs.~\ref{fig:fig4} and~\ref{fig:fig_ensemble}). Determining the universality classes of the waterbag content will be key to constructing a theory of universal equilibria (see, however, \cite{banik2025collisionless} for an alternative approach using quasilinear theory). Beyond developing a theory for the evolution of $\rho(\eta)$, applying our framework to 3D systems will require regularizing the Lynden-Bell equilibria, which have infinite mass at finite $\beta$ \cite{lynden1967statistical}; this can be done, e.g., by confining the system to a finite spatial domain or by imposing an energy cutoff \cite{lynden1967statistical, lynden1968gravo, chavanis1998degenerate}.
	 
	 The dissipation timescale \eqref{tau_dissp}, which is critical to our proposed picture of relaxation, arises from gravitational turbulence below the Jeans scale. We have derived the universal scaling laws \eqref{F_ks_solved} and \eqref{gk} for the fluctuation spectra of the DF and gravitational field governing this turbulence. Chaotic mixing of the DF in phase space underlies this gravitational turbulence. In particular, it is the exponential separation of particle trajectories in phase space that makes the dissipation timescale logarithmic in $N$ (cf. power-law-in-$N$ scalings that arise in non-turbulent systems \cite{su1968collisional, farias2018entropy, e2019discreteness2}). It has long been thought that chaotic mixing is intimately intertwined with violent relaxation \cite{merritt1996chaos,kandrup2003transient, kandrup2003chaos, de2011effectiveness}: our theory makes this connection~specific.
	
Self-similar gravitational collapse (e.g., \cite{bertschinger1985self,binney2004discreteness,colombi2014vlasov,parichha2025dark, sartorello2025chaos}) is also commonly regarded as a form of violent relaxation. While such collapses do involve time-evolving potentials, they are highly symmetric and laminar. We therefore do not expect our results to apply to such idealized collapses: their gravitational fields lack the irregular spatiotemporal structure that enables chaotic phase-space mixing of the~DF.
	
	We validated our turbulence theory with 1D $N$-body simulations (see also \cite{ginat2025gravitational}); testing it in 3D is an important challenge. Under the assumption that baryonic gas can be neglected, our theory should describe fluctuations of stars and dark matter below the Jeans scale during violent relaxation. Thus, one might speculate that our theory should describe the formation and merging of dwarf galaxies and elliptical galaxies \cite{mo2010galaxy}, which contain relatively little gas. It is an interesting task for future research to extend this theory to situations in which gas dynamics cannot be neglected and to determine how baryons affect the turbulent spectra derived here.
	
Outside the violent-relaxation paradigm, virialized systems may also exhibit gravitational turbulence, e.g., turbulent fluctuations in a disk galaxy could be excited by transient spiral waves or molecular clouds \cite{binney1988diffusion}. However, depending on the smallness of the fluctuation amplitude $\delta \rho_{m, r_{\mathrm{J}}}/\rho_{m, 0}$ in competition with the largeness of $N_{\mathrm{J}}$ in \eqref{k_noise}, the width of the inertial range may not be substantial. If such a turbulence does exist, it may be directly measurable with data from the Gaia spacecraft \cite{prusti2016gaia}.

	\section*{Methods}
	
	\subsection*{Details of numerical experiments} \label{appA}
	
\subsubsection*{$N$-body simulations}

The simulations in this paper were performed with the open-source $N$-body code \texttt{wendy}~\cite{wendy}; a description of the code is given in Appendix~A of \cite{bennett2021did}.

To construct the initial conditions of our simulations, we draw the centers of ten waterbags at random from a rectangular phase-space domain whose sides are ten times the waterbag width in each of $x$ and $v$. To prevent overlaps, we reject any center that lies within $2.5$ waterbag widths of an existing one in both $x$ and $v$. Each waterbag is filled uniformly with $N/10$ particles. We then shift the system to its center-of-mass frame and rescale particle velocities and positions, which sets $K$ and $U$, respectively, to reach a target total energy~$E = K + U$ and initial virial ratio $2K/U = 0.1$. We repeat this procedure with 100 different random seeds to produce an ensemble of 100 realizations for each of $N = 10^5$, $3\times 10^5$, $10^6$, $3\times 10^6$, and $10^7$. The ensembles for different $N$ use the same set of waterbag centers, so corresponding members of the ensembles differ only in the number of particles.

\subsubsection*{Gauge of the gravitational potential}

Several quantities in this paper---the total energy $E$ (and hence our unit of length $L = E/G M^2$), the virial ratio $2K/U$, and the single-particle energy per unit mass~$\epsilon$---depend on the gauge of $\Phi$, i.e., on its additive constant, which does not affect the dynamics. In 3D, the gauge is conventionally fixed by setting $\Phi(\infty) = 0$, but this is not possible in 1D, where $\Phi$ grows as $2\pi G M|x|$ at large $|x|$. The Green's function of the 1D Poisson equation~\eqref{Poisson} is $2\pi G|x - x'|$ up to an additive constant; we fix the gauge by setting this constant to zero, so that
\begin{equation} \label{Phi_1D}
	\Phi(x) = 2\pi G \int \mathrm{d}x'\,\rho_m(x')\,|x - x'|.
\end{equation}
This is the only gauge in which the virial theorem takes the simple form $2K = U$ (see, e.g.,~\cite{levin2014nonequilibrium}).
	
		\subsubsection*{Phase-space binning}
	\label{binning_sect}
	
	For each diagnostic that depends on a reconstructed DF, we first bin the particle coordinates $(x,v)$ onto a phase-space grid to obtain $f(x,v)$. Here we describe how we choose this grid.
	
	To calculate the phase-space spectra, we simply choose a resolution that is sufficiently fine to capture the dissipation cutoff $(k_{\mathrm{diss}}, s_{\mathrm{diss}})$ between the turbulence and noise spectra.
	
	To calculate the waterbag content \eqref{rho_eta_vlasov}, which is subsequently used to calculate the Lynden-Bell equilibria \eqref{LB_general}, we follow the prescription stated in the main text that $f(x,v)$ should be defined via a coarse-graining at the transition scale $(r_{\mathrm{diss}}, u_{\mathrm{diss}})$. We determine $k_{\mathrm{diss}}$ and $s_{\mathrm{diss}}$ empirically at each time by computing the wavenumbers where the ensemble-averaged spectra of the mass density~$\rho_m$ and of the velocity-space density~$\int \mathrm{d}x\, f$, respectively [i.e., the~$s = 0$ and $k = 0$ cuts of $P_f(k, s)$], fall below twice the amplitude of the noise floor \eqref{noise_spec} (the factor of two is a conservative margin to account for statistical fluctuations of the measured spectra about the noise floor). We then set the spatial grid spacing to $\Delta x = \pi/k_{\mathrm{diss}}$ and the velocity grid spacing to $\Delta v = \pi/s_{\mathrm{diss}}$, so that $k_{\mathrm{diss}}$ and $s_{\mathrm{diss}}$ are the Nyquist wavenumbers of the grid. Throughout the evolution, the resulting grid spacings satisfy $\Delta v \sim \tau_{\rm dyn}^{-1} \Delta x$, consistent with the turbulence being critically balanced.
	
	To calculate each ground-state distribution, we restack the corresponding $f(x,v)$ using the same binning procedure as for the waterbag content.
	
	For the contour plots of the DF in Fig.~\ref{fig:fig1}, we bin the particles on grids with an average of eight particles per phase-space cell. For the scatter plots of the coarse-grained DF in Figs.~\ref{fig:fig3} and~\ref{fig:fig_ensemble}, we use a fixed grid with $\Delta x = 0.016\,L$ and $\Delta v = 0.016\,v_0$. 
	
	We found that our results were largely insensitive to varying these resolutions (and the noise-floor margin) by factors of order unity.

	\subsubsection*{Computation of Lynden-Bell equilibria}
	
	Here, we describe the numerical computation of the Lynden-Bell equilibria shown in Figs.~\ref{fig:fig3} and~\ref{fig:fig_ensemble}. The inputs to~\eqref{LB_general} from the $N$-body simulation are the system's energy~$E$, which is fixed by the initial condition, and its waterbag content \eqref{rho_eta_vlasov}, which is recomputed at each instant in time. We obtain the waterbag content $\rho(\eta) = -\mathrm{d} \Gamma/\mathrm{d} \eta$ from its integral, $\Gamma(\eta) \equiv \int_{\eta}^{\infty} \mathrm{d} \eta'\,\rho(\eta') = \iint \mathrm{d}x\,\mathrm{d}v\, H(f-\eta)$. The function $\Gamma(\eta)$ is computed from $f(x,v)$ binned on an $(x,v)$-grid whose resolution is set by the procedure described above. We evaluate $\Gamma(\eta)$ on a uniformly spaced $\eta$ grid with spacing $\Delta\eta = \eta_{\mathrm{min}}$, running from $\eta_{\mathrm{min}} = m/\Delta x\, \Delta v$, corresponding to one particle per phase-space cell, up to~$\eta_{\mathrm{max}}$, the maximum value of $f$ at that time.
	
	The Lynden-Bell equilibria \eqref{LB_general} are computed via three nested iteration loops. The outermost loop enforces self-consistency of the potential via Poisson's equation, \eqref{Poisson_LB}, the middle loop increments $\beta$ to satisfy the energy constraint \eqref{energy}, and the inner loop iterates the chemical potentials, encoded in the ``fugacities'' $F(\eta) = \mathrm{e}^{\beta\eta \mu(\eta)}$, to satisfy the waterbag-content constraint \eqref{rho_eta}.
	
	The phase-space grid is taken sufficiently large for the neglected tail of the distribution to be numerically irrelevant. To speed up the iterations, phase-space integrals are reduced to energy integrals weighted by the density of states \eqref{density_states}.
	
	\subsubsection*{Computation of ground states} \label{gs_section}
	
	In the main text, we computed the ground state by ``restacking,'' i.e., by energy-ordering the DF's level sets (waterbags): the largest $\eta$ was assigned to the lowest $\epsilon(x, v) = v^2/2+ \Phi(x)$, with $\eta$ decreasing monotonically with $\epsilon$. Restacking preserves phase-space volume but not, in general, the self-consistent potential $\Phi$. We therefore recompute $\Phi(x)$ from the restacked DF, update $\epsilon$, and iterate until convergence. The resulting fixed point is a phase-space-volume-preserving DF that decreases monotonically with energy; assuming that the ground state is unique (see the section ``From initial state to ground state'' in the main text), it minimizes the total energy~\cite{gardner1963bound,dodin2005variational,helander2017available,wiechen1988relaxation,aly1989lowest}.

	\subsubsection*{Computation of dissipation timescale} \label{tau_dissp_appendix}
	
To determine $\tau_{\mathrm{diss}}$ given by \eqref{tau_dissp_from_C2} and shown in Fig.~\ref{fig:fig7}, we first compute, at each time, the amount of $C_2$ \eqref{C2} contained in modes above the noise floor:
\begin{equation} 
	\overline{C_2^{\,>\mathrm{noise}}} = \frac{V}{(2 \pi)^2} \iint_{P_f \geq 2 \rho_{m,V}^2/N} \mathrm{d} k \, \mathrm{d} s \, P_{f}(k, s),
\end{equation}
where $\rho_{m,V}^2/N$ is the amplitude of the noise floor [see~\eqref{noise_spec} and Figs.~\ref{fig:fig5} and~\ref{fig:fig6}], and the factor of two is the same margin as used for the phase-space binning. This is a proxy for the $C_2$ stored below the dissipation cutoff $(k_{\mathrm{diss}}, s_{\mathrm{diss}})$, with the region of $(k,s)$ space determined directly from the data. We then define $\tau_{\mathrm{diss}}$ as the first time at which $\overline{C_2^{\,>\mathrm{noise}}}$ falls to $95\%$ of its value at $t = 0$.
	
	\subsection*{Scaling theory for sub-Jeans cascade} \label{appB}
	
		\begin{figure*}[t]
		\centering
		\includegraphics[width=\textwidth]{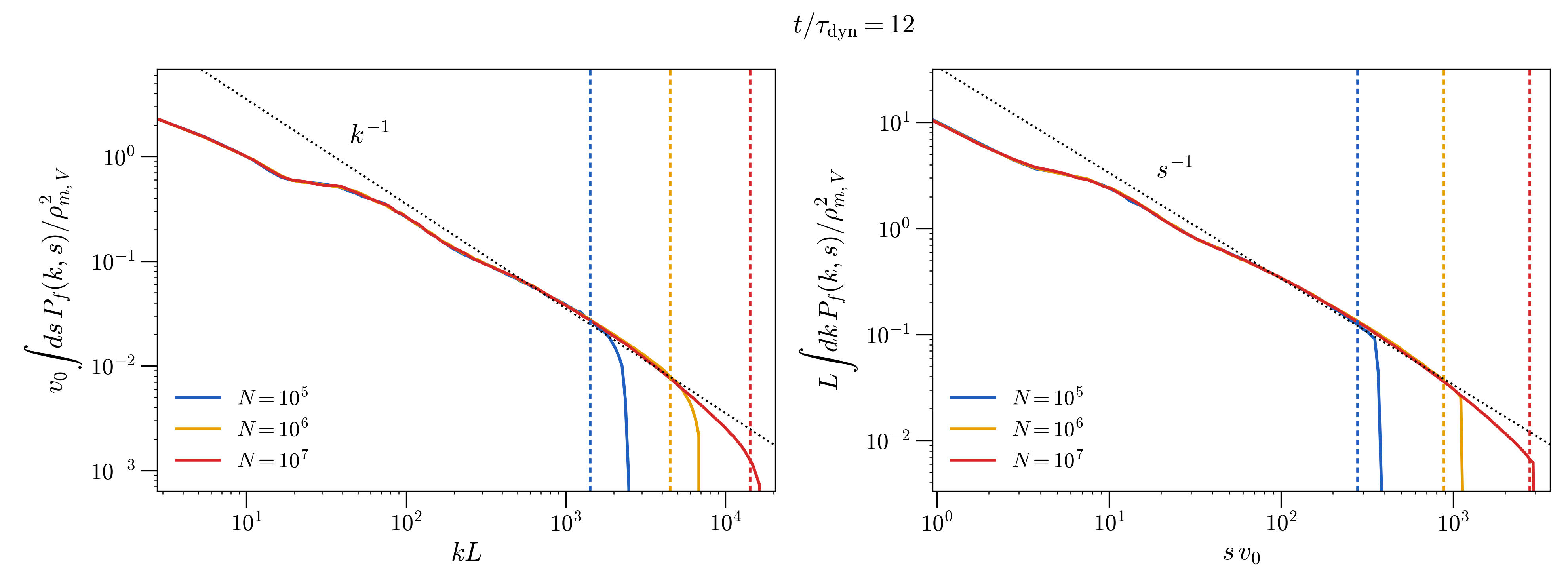}
		\caption{\label{fig:fig8} 1D spectra of $C_2$ at $t = 12\,\tau_{\rm dyn}$, ensemble-averaged over 100 simulations for each of $N = 10^5$ (blue), $10^6$ (orange), and $10^7$ (red). Left panel: the spectrum in $k$, obtained by integrating the 2D spectrum $P_f(k,s)$ (shown for $N = 10^7$ in Fig.~\ref{fig:fig6}) over $s$. Right panel: the spectrum in $s$, obtained by integrating $P_f(k,s)$ over $k$. To prevent the noise floor from contaminating the 1D spectra, the integral over $s$ ($k$) is truncated at $s_{\mathrm{diss}}$ ($k_{\mathrm{diss}}$), estimated for each $N$ as the first wavenumber at which the $C_2$ spectrum falls below two times the noise floor \eqref{noise_spec}. In each panel, the red vertical line marks the wavenumber at which the $N=10^7$ spectrum drops off; the other two vertical lines are obtained from it using the scaling \eqref{k_noise} (and, for $s$, by critical balance, $s_{\mathrm{diss}} \propto k_{\mathrm{diss}}$).
		}
		
	\end{figure*}
	
	Here we present a phenomenological scaling theory for the phase-space spectrum \eqref{F_ks_solved}, following the plasma theory of \cite{nastac2025universal}.
	
	\subsubsection*{$C_2$ budget}
	
	We can decompose $C_2$ into the parts containing modes above and below the Jeans scale: $\overline{C_2} = \overline{C_2^{<k_{\mathrm{J}}}} + \overline{C_2^{>k_{\mathrm{J}}}}$, where $k_{\mathrm{J}} \sim r_{\mathrm{J}}^{-1}$, $ \overline{C_2^{<k_{\mathrm{J}}}} = V/(2 \pi)^{2d} \iint_{|\mathbf{k}| < k_{\mathrm{J}}} \mathrm{d} \mathbf{k} \mathrm{d} \mathbf{s} \, P_{f}(\mathbf{k}, \mathbf{s})$, and $ \overline{C_2^{>k_{\mathrm{J}}}} = V/(2 \pi)^{2d} \iint_{|\mathbf{k}| > k_{\mathrm{J}}} \mathrm{d} \mathbf{k} \mathrm{d} \mathbf{s} \, P_{f}(\mathbf{k}, \mathbf{s})$. The $C_2$ conservation yields the following budget equation:
	\begin{equation} \label{C2_budget}
		\frac{\mathrm{d}}{\mathrm{d} t}\overline{C_2^{>k_{\mathrm{J}}}} = \varepsilon-\mathcal{D},
	\end{equation}
	where $\varepsilon = -\mathrm{d} \overline{C_2^{<k_{\mathrm{J}}}}/ \mathrm{d} t$, and $\mathcal{D} \geq 0$ represents the dissipation of $C_2$ by finite-$N$ effects. We assume that the system reaches a quasi-steady state in which $\mathrm{d} \overline{C_2^{>k_{\mathrm{J}}}}/\mathrm{d} t \simeq 0$, so that \eqref{C2_budget} implies $\varepsilon = \mathcal{D}$. We posit that $C_2$ is injected into sub-Jeans scales predominantly at the Jeans scale and that the dissipation occurs at some small length scale $r_{\mathrm{diss}} \ll r_{\mathrm{J}}$ (estimated in the main text). As we argue in the main text, the velocity-space scale of $C_2$ injection at the Jeans scale is $v_{\mathrm{th}}$, and the dissipation scale is $u_{\mathrm{diss}} \ll v_{\mathrm{th}}$. To bridge these disparate scales in phase space, we assume that there will be an ``inertial range'' of scales through which the flux $\varepsilon$ passes, unaffected directly by injection or dissipation. 
	
	\subsubsection*{Turbulent cascade}
	
	While the DF may be inhomogeneous and anisotropic at large scales, we assume that the DF is approximately statistically homogeneous and isotropic in the inertial range. We also ignore any dependence of the spectrum on the angle $\theta = \cos^{-1} (\mathbf{k} \cdot \mathbf{s}/k s)$. The latter assumption is not exactly valid: in order for there to exist a constant-flux cascade in phase space, there must be an asymmetry between the parts of the spectrum where $\mathbf{k} \cdot \mathbf{s} > 0$ and $\mathbf{k} \cdot \mathbf{s} < 0$. This is shown explicitly in Appendix B of \cite{nastac2025universal} for a Vlasov--Poisson plasma, but their calculation applies equally well to gravitational systems. We assume that, in each of the asymptotic limits $k \gg \gamma s$ and $k \ll \gamma s$, the spectrum is a power law to leading order, with its $\theta$-dependence entering only at higher order in $\gamma s/k$ or $k/\gamma s$, respectively. Our analysis is therefore simplified to predicting the fluctuation spectrum shell-averaged in both $\mathbf{k}$ and $\mathbf{s}$, which we denote by $\langle P_{f} \rangle_{k,s}$.
	
	We propose that the system undergoes a constant-flux cascade of $C_2$ in both position and velocity space:
	\begin{equation} \label{constant_flux}
		\frac{v_{\mathrm{th}}^d \delta \! f_{r}^2}{\tau_{\mathrm{c}}} \sim \varepsilon, \quad \frac{v_{\mathrm{th}}^d \delta \! f_{u}^2}{\tau_{\mathrm{c}}} \sim \varepsilon,
	\end{equation}
	where $\delta \! f_{r}$ and $\delta \! f_{u}$ are typical spatial and velocity increments of $f$ across the scales $r$ and $u$, respectively, and $\tau_{\mathrm{c}}$ is the cascade time, which is the typical time that it takes for $C_2$ to be transferred across phase-space scales~$(r,u)$.
	
	\subsubsection*{Critical balance}
	
	To determine $\tau_{\mathrm{c}}$, note that there are two processes by which fine-scale structure can be generated in \eqref{Vlasov}: linear phase mixing due to ballistic particle streaming, which produces fine structure in velocity space on the timescale $\tau_{\mathrm{lin}} \sim r/u$, and nonlinear mixing due to particle acceleration by gravitational fields, which produces fine-scale structure in position space on the timescale $\tau_{\mathrm{nl}} \sim u/ \delta g_r$, where $\delta g_r$ is the typical gravitational field increment across scale $r$. We assume (and verify \textit{a posteriori}) that the gravitational field is spatially smooth, so that $\delta g_r \sim \gamma^2 r$, where $\gamma$ is the gravitational ``shearing'' rate.
	
	We argue that $\tau_{\mathrm{c}}$ is given by the critical balance between the linear and nonlinear times \cite{nastac2023phase,nastac2025universal,ginat2025gravitational}, by analogy with the critical-balance conjecture in magnetohydrodynamic \cite{goldreich1995toward,schekochihin2022mhd} and plasma-kinetic \cite{nastac2023phase,nastac2025universal} turbulence. Any initial fluctuation with $\tau_{\mathrm{lin}} < \tau_{\mathrm{nl}}$ will be linearly phase-mixed to smaller $u$, thereby reducing~$\tau_{\mathrm{nl}}$. Likewise, one with $\tau_{\mathrm{nl}} < \tau_{\mathrm{lin}}$ will be nonlinearly phase-mixed to smaller $r$, lowering $\tau_{\mathrm{lin}}$. Hence,~$\tau_{\mathrm{c}} \sim \tau_{\mathrm{lin}} \sim \tau_{\mathrm{nl}} \implies \tau_{\mathrm{c}} \sim \gamma^{-1}$. This means that the DF undergoes mixing concurrently in position and velocity space. This is consistent with the exponential separation of particle trajectories, leading to simultaneous stretching of structures in both spaces, as discussed in the main text.

	\subsubsection*{Phase-space spectra}
	
Putting $\tau_{\mathrm{c}} \sim \gamma^{-1}$ into \eqref{constant_flux} gives us the 1D spectra of~$C_2$:
	\begin{align} \label{1D_spectra}
		\int \mathrm{d} s \, \langle P_{f} \rangle_{k,s} \sim \varepsilon \gamma^{-1} k^{-1}, \, \int \mathrm{d} k \, \langle P_{f} \rangle_{k,s} \sim \varepsilon \gamma^{-1} s^{-1}.
	\end{align}
	These scalings are the gravitational Vlasov--Poisson analogues of the Batchelor $k^{-1}$ spectrum for a passive scalar advected by a smooth flow \cite{batchelor1959small}. We confirm them numerically in Fig.~\ref{fig:fig8}.
	
	To determine the 2D spectrum of $C_2$ that satisfies \eqref{1D_spectra}, we assume that it is a product of power laws, with a break at the critical-balance line $k \sim \gamma s$. The only 2D spectrum consistent with \eqref{1D_spectra} is \eqref{F_ks_solved} (cf. \cite{nastac2025universal}). The scalings $k^{d-1}, s^{d-1}$ (which come from the wavenumber Jacobians in $d$ dimensions) arise from assuming that the non-shell-averaged spectrum does not depend on $k$ or $s$ in the long wavelength limits. The justification for this assumption is as follows. By critical balance, the position-space correlation scale of the turbulence at a given velocity-space scale $u$ is simply $r \sim \gamma^{-1} u$. The spectrum at long wavelengths measures correlations between points in phase space that are separated by more than one correlation scale.
	Such points are effectively uncorrelated, so the spectrum at these wavenumbers is that of a white noise in position (velocity) space and is, therefore, independent of $k$ ($s$). Then, the high-($k$, $s$) scalings in \eqref{F_ks_solved} are fixed by enforcing that taking the integral over $k$ and $s$ of \eqref{F_ks_solved} gives \eqref{1D_spectra}. These are the only possible scalings consistent with a critically balanced, constant-flux phase-space cascade of $C_2$. This was demonstrated in \cite{nastac2025universal} (Appendix A) for electrostatic Vlasov--Poisson plasma turbulence below the Debye scale, but the same calculation applies to gravitational systems below the Jeans scale.
	
	We stress the distinction between \eqref{F_ks_solved} and the super-Jeans-scale phase-space spectrum derived by \cite{ginat2025gravitational} (see their Eq.~41). In both regimes, the gravitational field is spatially smooth, but above the Jeans scale, the system is unstable to gravitational collapse \cite{jeans1902stability}. This introduces an injection of $C_2$ at every (Jeans-unstable) scale, causing the flux of $C_2$ to accumulate $\propto k^d$ in $k$ space (and, by critical balance, $\propto s^d$ in $s$ space). The result is a fluctuation spectrum with shallower power-laws than the sub-Jeans-scale spectrum, viz., $\langle P_{\mathbf{g}}\rangle_{k} \propto k^{-3}$ [cf. \eqref{gk}]. Below the Jeans scale, the system is no longer unstable to gravitational collapse, and the flux becomes constant, resulting in \eqref{F_ks_solved}.

	\section*{Acknowledgements}
	
	We are thankful to T. Adkins, U. Banik, J. Binney, F. van den Bosch, S. Colombi, S. Flores, J.-B. Fouvry, C. Hamilton, D. Hosking, M. Kunz, H. Latter, J. Magorrian, C. Pichon, W. Sengupta, J. Squire, A. L. Varri, E. Vasiliev, and M. Weinberg for helpful discussions related to this work. M.L.N. was supported by a Clarendon Scholarship. Y.B.G.'s research was supported by the Royal Commission for the Exhibition of 1851, and partly by a Leverhulme Trust International Professorship Grant to S. L. Sondhi (No.~LIP-2020-014). M.L.N., Y.B.G., and A.A.S. were supported in part by a Simons Investigator Award to A.A.S. from the Simons Foundation. R.J.E. was supported by the Simons Foundation grant MP-SCMPS-00001470. M.L.N., Y.B.G., and R.J.E. were also supported in part by the grants GALBAR ANR-25-CE31-4684 and BEYOND-BL ANR-25-CE57-2626, as well as the CNRS through the MITI interdisciplinary programs. Y.B.G.’s work was performed in part at the Aspen Center for Physics, which is supported by a grant from the Simons Foundation (1161654, Troyer). M.L.N., R.J.E., M.B., and A.A.S. are also grateful to the Wolfgang Pauli Institute, Vienna, for its hospitality on multiple occasions. The simulations performed for this paper were done on the Stellar cluster at Princeton University. Claude AI (Opus 5.5) was used to refine and adapt author-provided code for data analysis and visualization, as well as to assist with proof-reading the manuscript.
	
	\bibliography{paper}
	
\end{document}